\documentclass[12pt,aps,prd,reprint,sort&compress, superscriptaddress, showpacs, nofootinbib, preprintnumbers]{revtex4-1}

\usepackage{graphicx}
\usepackage{amsmath}
\usepackage{amssymb}
\usepackage{dsfont}
\usepackage{amsfonts}
\usepackage[UKenglish]{babel}
\usepackage{hyperref}
\usepackage{nicefrac}
\usepackage{verbatim}
\usepackage{bbm}
\usepackage{color}
\usepackage{multirow}
\usepackage{subcaption}
\usepackage{cleveref}

\newcommand{\E}{\mathrm{e}}

\newcommand{\tab}[1]{Tab.~{\ref{#1}}}

\newcommand{\cg}{\gamma_5}

\begin{document}

\title{Emergence of a new $SU(4)$ symmetry in the baryon spectrum}
 
\author{M.~Denissenya}
\email{mikhail.denissenya@uni-graz.at}
\author{L.~Ya.~Glozman}
\email{leonid.glozman@uni-graz.at}
\author{M.~Pak}
\email{markus.pak@uni-graz.at}
\affiliation{Institut f\"ur Physik, FB Theoretische Physik, Universit\"at Graz, Universit\"atsplatz 5,
8010 Graz, Austria}

\begin{abstract}
Recently a large degeneracy of $J=1$ mesons, that
is larger than the $SU(2)_L \times SU(2)_R \times U(1)_A$ symmetry of the QCD
Lagrangian, has been discovered upon truncation of the near-zero modes from the valence quark
propagators.  It has been found that this degeneracy represents the
$SU(4)$ group that includes the chiral rotations as well as the mixing of 
left- and right-handed quarks. This symmetry group turnes out to be a symmetry
of confinement in QCD. Consequently, one expects that the same symmetry 
should persist upon the near-zero mode removal in other hadron sectors as well. It has been shown that indeed
the $J=2$ mesons follow the same symmetry pattern upon the low-lying mode
elimination. Here we demonstrate  the  $SU(4)$ symmetry of baryons once 
the near-zero modes are removed from the quark propagators.
We also show a degeneracy of states belonging to different irreducible representations of $SU(4)$. This implies a larger symmetry, that includes $SU(4)$ as a subgroup.

\end{abstract}

\maketitle

\section{Introduction}
In recent two-flavor lattice simulations with the manifestly chiral-invariant Overlap Dirac operator a large degeneracy of $J=1$ mesons has been discovered upon
truncation of the near-zero modes from the valence quark propagators
\cite{Denissenya:2014ywa,Denissenya:2014poa}. This degeneracy turned out to
be larger than the $SU(2)_L \times SU(2)_R \times U(1)_A$ symmetry of the QCD
Lagrangian. 

In Ref.~\cite{Glozman:2014mka} the symmetry group that drives
this degeneracy has been reconstructed, that is $SU(4) \supset SU(2)_L \times SU(2)_R \times U(1)_A$. This symmetry "rotates" the fundamental vector
$(u_L, u_R, d_L, d_R)^T$ and includes both the $SU(2)_L \times SU(2)_R$
chiral rotations as well the $SU(2)_U \times SU(2)_D$ rotations in the \textit{chiralspin} space that mix the left- and right-handed components of the quark fields. It has been shown that there are no magnetic interactions between quarks in the system after truncation and consequently a meson after
truncation represents
a dynamical quark-antiquark system connected by the electric field. Such a system has been interpreted as a dynamical QCD string and the $SU(4)$
symmetry has been identified to be a symmetry of confinement. 

This symmetry has
been studied in detail in Ref.  \cite{Glozman:2015qva}, where  transformation properties of different operators have been obtained. It has also been found
that this symmetry is hidden in the QCD Hamiltonian in Coulomb gauge, namely in the Coulombic interaction part. The confining charge-charge
part is a $SU(4)$-singlet and generates therefore a $SU(4)$-
symmetric spectrum. Interactions of quarks with the magnetic field
 explicitly break this symmetry of the confinement part and are also responsible for the $SU(2)_L \times SU(2)_R$ and  $U(1)_A$ breakings. 
 
 The appearance of the $SU(4)$ symmetry has also been demonstrated in $J=2$
 mesons, which can be found in Ref.~\cite{Denissenya:2015mqa}.
  
Here we study if $SU(4)$ also appears in the $N$ and $\Delta$ baryon spectrum after removing the quasi-zero modes. 
The elimination of the quasi-zero modes has been accomplished earlier
with Wilson-type fermions (Chirally Improved fermions) in the baryon sector in Ref.~\cite{Glozman:2012fj}, where the chiral restoration has been
studied. The question of the $SU(4)$ could not be addressed at that time, however.

The outline of the article is as follows: In Chapter \ref{Chapter-Chiral-Parity} we review the classification of baryon operators according to the 
irreducible representations of the parity-chiral group and show, which states should become mass degenerate if the $SU(2)_{CS}$ and higher $SU(4)$ symmetry is in the system. In Chapter
\ref{Chapter-Lattice-Setup} we shortly present the details of the lattice setup. In Chapter \ref{Chapter-Results} the eigenvalues
of the correlation matrix and effective masses are presented, which show 
$SU(2)_L \times SU(2)_R$,   $SU(2)_{CS}$ and $SU(4)$ symmetries upon
truncation of the near-zero modes.
In Chapter \ref{Conclusions} we summarize our findings. 

\section{$SU(2)_L \times SU(2)_R$, $U(1)_A$, $SU(2)_{CS}$ and
$SU(4)$  properties of baryon fields}
\label{Chapter-Chiral-Parity}

All possible two-flavor baryon fields can be classified according to
the $SU(2)_L \times SU(2)_R$ representations $r$ \cite{CJ}, see Table I.

\begin{table}
  \centering
  \renewcommand{\arraystretch}{1.5}
  \begin{tabular}{l c  }
   \text{Baryons} (isospin $I$) ~~~~~~~~~~~~~~& $r$\\
   $ N^\pm(I=\frac{1}{2})$                & $(\frac{1}{2},0)+(0,\frac{1}{2})$\\ 
  $N^\pm(I=\frac{1}{2}),\,\Delta^\pm(I=\frac{3}{2})$ & $(1,\frac{1}{2})+(\frac{1}{2},1)$\\
  $\Delta^\pm(I=\frac{3}{2})$ & $(\frac{3}{2},0)+(0,\frac{3}{2})$ 
  \end{tabular}
  \renewcommand{\arraystretch}{1}
\caption{Chiral multiplets for baryons with fixed total spin $J$, wirh $r$ being the index of the parity-chiral multiplet.}\label{tab:BR}
 \end{table}

Here we analyse standard nucleon and delta interpolators that are constructed
as
\begin{align}
\label{generalN}
 N^{(i)}_{\pm} &= \varepsilon_{abc} \mathcal{P}_{\pm} \Gamma_1^{(i)} u_a \left(u_b^T \Gamma^{(i)}_2 d_c - d^T_b \Gamma_2^{(i)} u_c \right) \;, \\ 
\label{generalD}
 \Delta^{(i)}_{\pm} &= \varepsilon_{abc}  \mathcal{P}_{\pm} \Gamma_1^{(i)} u_a \left(u_b^T \Gamma^{(i)}_2 u_c \right) \; , 
\end{align}
with the parity projector $\mathcal{P}_{\pm} =\frac{1}{2}(\mathds{1} \pm \gamma_0)$. 
The terms in the brackets we refer to as diquarks. For the spin-$\frac{3}{2}$ interpolators the Rarita-Schwinger projection is used.

For spin-$\frac{1}{2}$ nucleons three different Dirac structures $\chi^{i} = ( \Gamma_1^{i},  \Gamma_2^{i})$ are employed, see Table II. 
These three interpolators have distinct chiral transformation properties 
\cite{Brommel:2003jm}. All these fields are not connected to each other
through the $SU(2)_L \times SU(2)_R$ transformations.

For the spin-$\frac{1}{2}$ $\Delta$, the spin-$\frac{3}{2}$ nucleon and the 
spin-$\frac{3}{2}$ $\Delta$ only one type of interpolator for each
baryon  is analysed and taken into account, see Table \ref{tab:BI}. 

\begin{table}[thb]
\begin{center}
\renewcommand{\arraystretch}{1.5}
\begin{tabular}{c|cc|c}
 \hline
 \hline
 \multicolumn{1}{c|}{   $I,J^{P}$ } &$\Gamma^{(i)}_1$ & $\Gamma^{(i)}_2$ & $r$  
   
 \\
 \hline

 \multicolumn{1}{c|}{\multirow{3}{*}{$N(\frac{1}{2},\frac{1}{2}^\pm)$}} & $\mathds{1}$ & $C \cg $& \multicolumn{1}{c}{\multirow{1}{*}{$(\frac{1}{2},0)+(0,\frac{1}{2})$}}
 \\
 % \cline{2-4}
  & $ \cg $ & $C $& $(\frac{1}{2},0)+(0,\frac{1}{2})$\\

  & $i \mathds{1}$ & $C \cg \gamma_0 $ & $(1,\frac{1}{2})+(\frac{1}{2},1)$ \\
  $\Delta(\frac{3}{2},\frac{1}{2}^\pm)$ & $i \gamma_i\cg$  & $C \gamma_i$ &
\multicolumn{1}{c}{\multirow{1}{*}{$(1,\frac{1}{2})+(\frac{1}{2},1)$}}\\
\cline{2-4}
  $N(\frac{1}{2},\frac{3}{2}^\pm)$ & $i \cg$  & $C \gamma_i \cg$ &
\multicolumn{1}{c}{\multirow{1}{*}{$(1,\frac{1}{2})+(\frac{1}{2},1)$}}\\
  $\Delta(\frac{3}{2},\frac{3}{2}^\pm)$ & $i \mathds{1} $  & $C \gamma_i$ &
\multicolumn{1}{c}{\multirow{1}{*}{$(1,\frac{1}{2})+(\frac{1}{2},1)$}}\\
\hline
\hline
\end{tabular}
\renewcommand{\arraystretch}{1}
\end{center}
\caption{List of Dirac structures for the $N$ and $\Delta$ baryon fields
and respective  chiral representations $r$.}\label{tab:BI}
\end{table}

\begin{figure}
\centering
\includegraphics[angle=0,width=\linewidth]{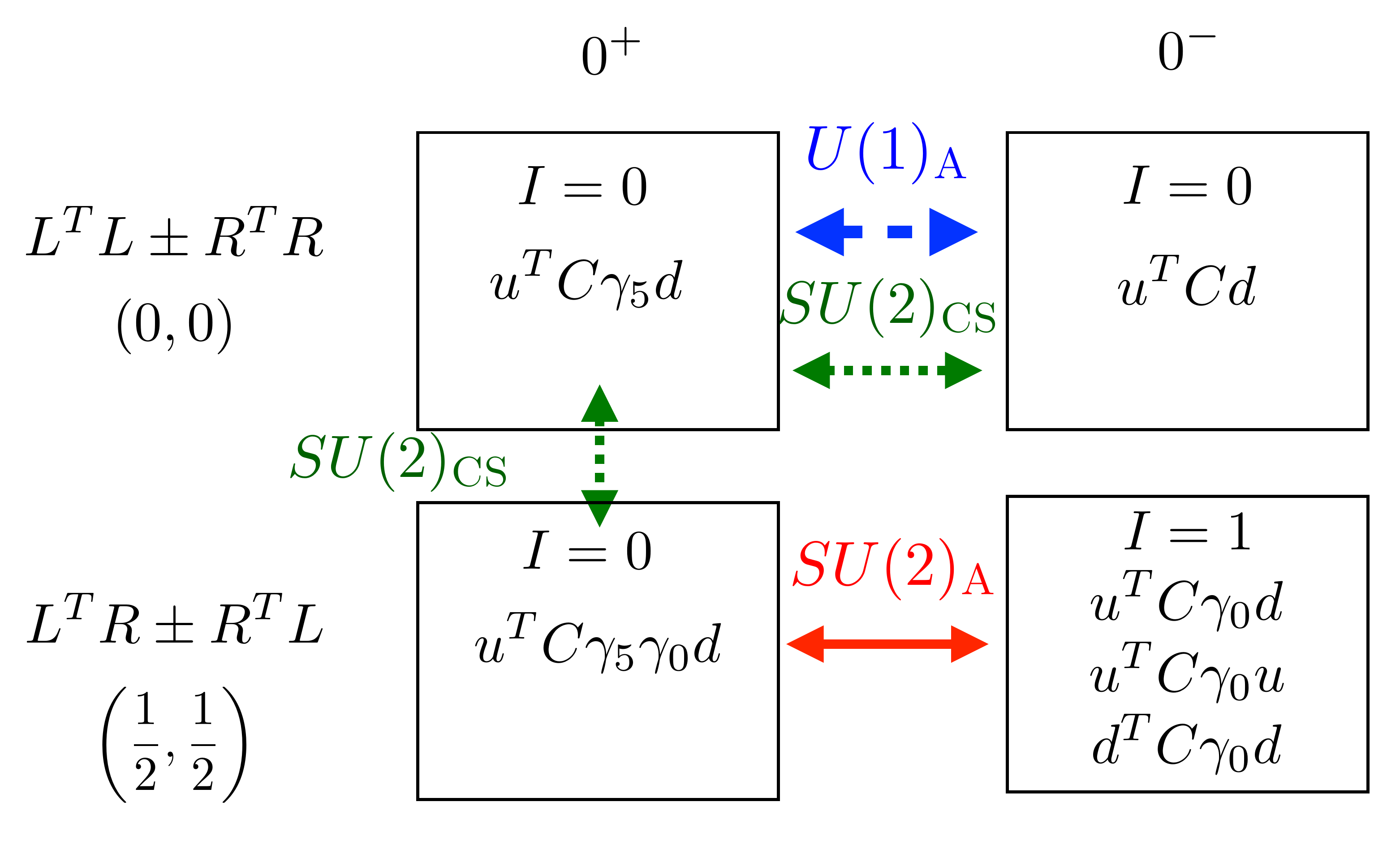}
\caption[Chiral-Parity group 1]{\sl In the left column the left/right-components of the $J=0$ diquarks are given. 
The interpolators for the $0^+$ and $0^-$ diquarks 
 are given in the second and third column. 
The color indices are suppressed, but can be read off from the general formula in eqs.~(\ref{generalN}) and (\ref{generalD}). 
The $SU(2)_A$ and $U(1)_A$ transformations are denoted by red and blue lines, respectively. 
The $SU(2)_{CS}$ transformation is given by a green line.
It combines interpolators with different left/right-components. 
The interpolator with Dirac structure $C \gamma_0$ is invariant with respect to $SU(2)_{CS}$. $SU(4)$ connects all different diquarks 
in this figure.}
 \label{Table1}
\end{figure}

Consider the following interpolator 
\begin{align}
\label{Interpolator1}
 \mathcal{O}_{N^{\pm}} = \varepsilon^{abc} \mathcal{P}_{\pm} u^a \left[ u^{b T} C \gamma_5 d^{c} - d^{b T} C \gamma_5 u^c \right] \; ,
\end{align}
which generates a spin-$\frac{1}{2}$ nucleon of positive and negative parity.
Here $C$ denotes the charge conjugation matrix. The scalar ($0^+$)
diquark interpolator $u^T C \gamma_5 d$  has
the following chiral content $u^T_L d_L + u^T_R d_R$. 
It is invariant with respect to the axial part of the 
$SU(2)_L \times SU(2)_R$ transformations (in the following we abbreviate these transformations as $SU(2)_A$), i.e. it belongs to a singlet $r=(0,0)$
representation.  Therefore, via 
$SU(2)_A$ the interpolator (\ref{Interpolator1})  mixes only with
\begin{align}
 \mathcal{O}_{N^{\mp}} &= \varepsilon^{abc} \mathcal{P}_{\pm} \gamma_5 u^a \left[ u^{b T} C \gamma_5 d^{c} - d^{b T} C \gamma_5 u^c \right] \; , \\
 \mathcal{O}_{N^{\mp}} &= \varepsilon^{abc} \mathcal{P}_{\pm} \gamma_5 d^a \left[ u^{b T} C \gamma_5 d^{c} - d^{b T} C \gamma_5 u^c \right] \; ,
\end{align}
which create nucleons with opposite parity. 
The interpolator (\ref{Interpolator1}) falls into the $(1/2,0) \oplus (0,1/2)$ representation and combines nucleons of positive and negative parity. 

Under $U(1)_A$ the diquark $u^T C \gamma_5 d$ mixes with the diquark
$u^T C d$. Therefore, the interpolator (\ref{Interpolator1}) gets mixed
with the interpolator from the second line of Table \ref{tab:BI}. The $SU(2)_A$, $U(1)_A$
and $SU(2)_{CS}$ connections of different $J=0$ diquarks are illustrated in 
Fig.~\ref{Table1}.

Now we look at the following interpolator 
\begin{align}
\label{Interpolator2}
 \mathcal{O}_{N^{\pm}} = i \varepsilon^{abc} \mathcal{P}_{\pm} u^a \left[ u^{b T} C \gamma_5 \gamma_0 d^{c} - d^{b T} C \gamma_5 \gamma_0 u^c \right] \; ,
\end{align}
which also generates a spin-$\frac{1}{2}$ nucleon of both parities. The only difference to the field (\ref{Interpolator1}) is that the diquark has an 
additional $\gamma_0$ structure. The diquark  $u^T C \gamma_5 \gamma_0 d$ is  a scalar ($0^{+}$), 
but has a different chiral content, namely $u^T_L d_R + d^T_L u_R$. 
It belongs to a $(1/2,1/2)$ chiral representation.
Via $SU(2)_A$ it mixes with diquarks 
\begin{align}
 u^{T} C \gamma_0 d \; ,~~ u^{T} C \gamma_0 u \; ,~~ d^{T} C \gamma_0 d\; , 
\end{align}
that form the $I=1$ triplet.
With the latter diquarks one can construct an  interpolator 
for deltas with spin $J=1/2$, that is distinct from
 the one in Table II.

Therefore, the interpolator (\ref{Interpolator2}) belongs to the $(1, 1/2) \oplus (1/2,1)$ representation of the parity-chiral group.

Under $U(1)_A$ the diquarks  $u^T C \gamma_5 \gamma_0 d$, $u^T C \gamma_0 d$, $u^T C \gamma_0 u$, $d^T C \gamma_0 d$ are selfdual. 

In the $J=3/2$ interpolators that we use (two last lines in Table \ref{tab:BI})
the $J=1$ diquarks are connected to each other via the $SU(2)_A$ transformations and both $N(\frac{1}{2},{\frac{3}{2}}^{\pm})$ and
$\Delta(\frac{3}{2},{\frac{3}{2}}^{\pm})$ interpolators form
a $(1,\frac{1}{2})+(\frac{1}{2},1)$ - 12-plet of the parity-chiral group.

 \begin{figure*}[t!]
    \centering 
      \begin{subfigure}[b]{0.45\textwidth}
        \centering
        \includegraphics[scale=0.5]{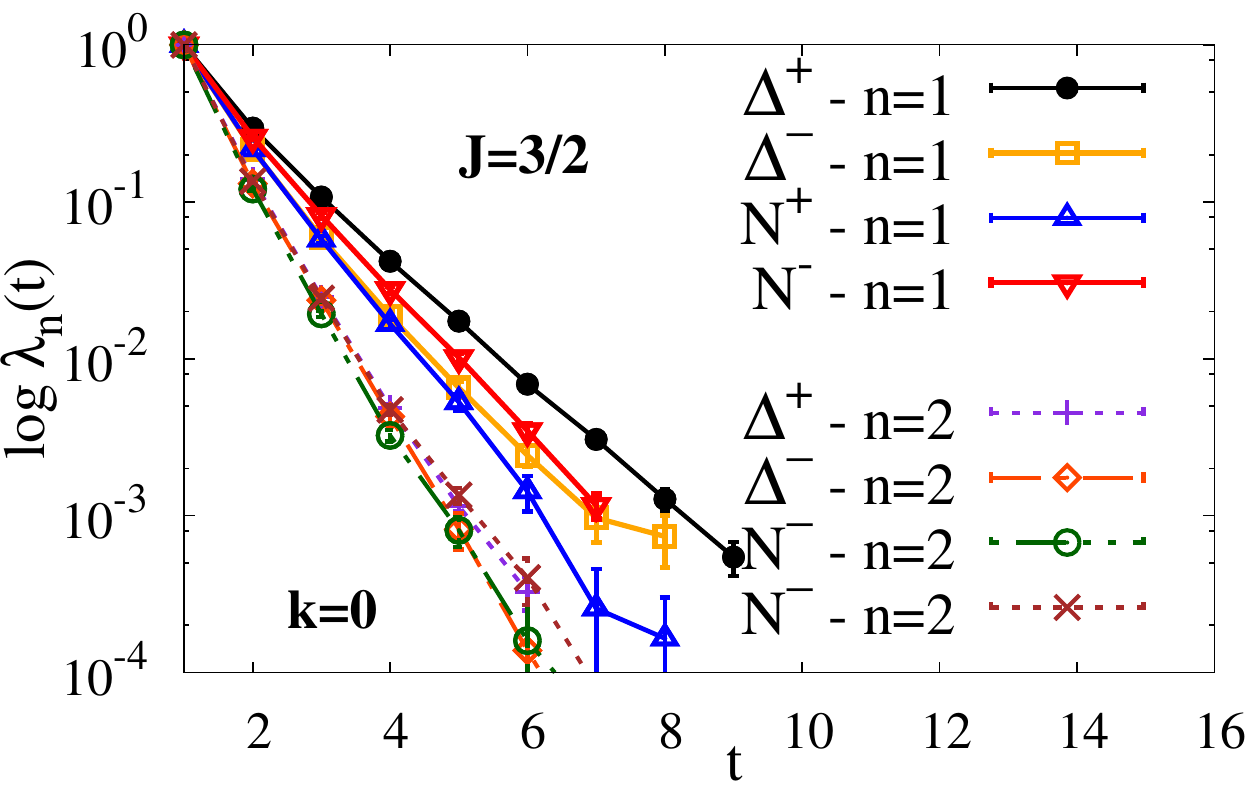}
        \caption{}
      \end{subfigure}
       \hspace*{-32pt}
      \begin{subfigure}[b]{0.45\textwidth}
          \centering
          \includegraphics[scale=0.5]{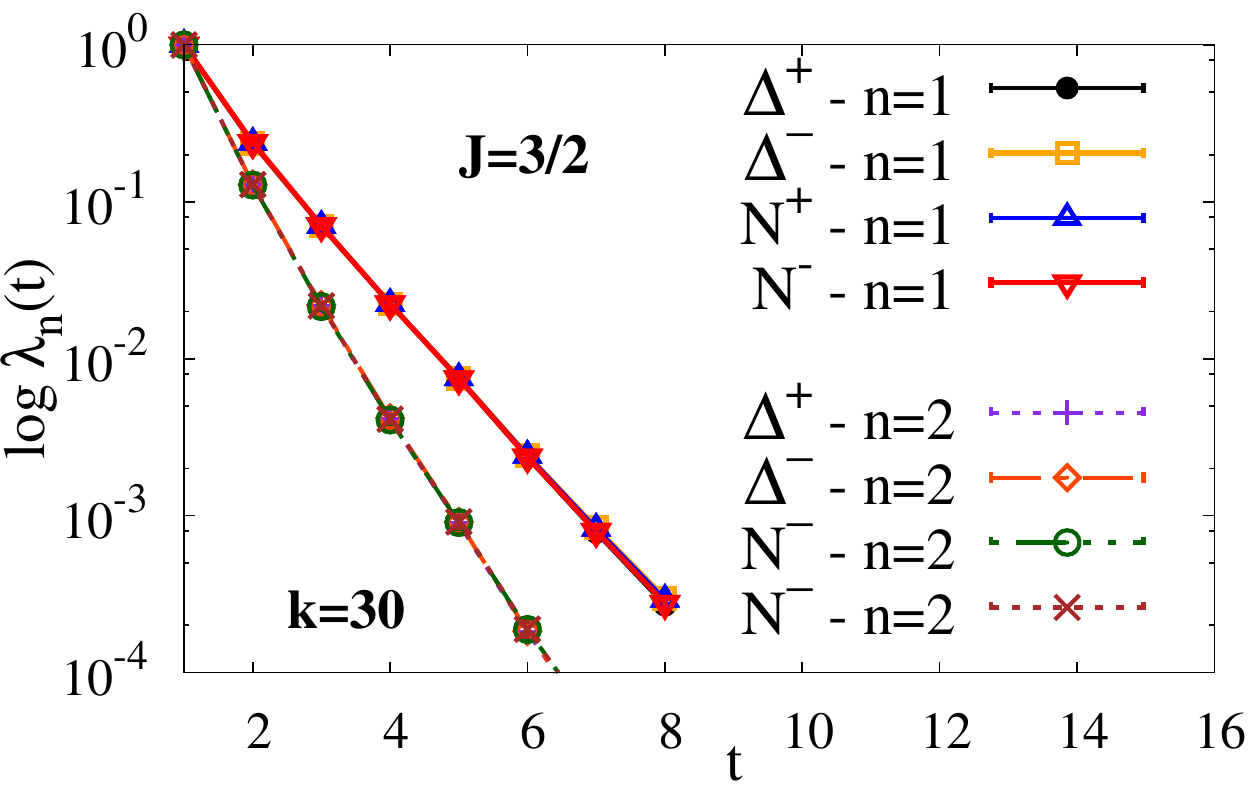}
          \caption{}
      \end{subfigure}
     \hspace*{24pt} \hfill\\ 
      \caption{Eigenvalues of $J=\frac{3}{2}^{\pm}$ baryons: (a) full case ($k=0$);
     (b)  after excluding $k=30$ quasi-zero modes.
}\label{fig:BB32ev}
   \end{figure*}

\begin{figure*}[htb]
    \centering 
      \begin{subfigure}[b]{0.45\textwidth}
        \centering
        \includegraphics[scale=0.5]{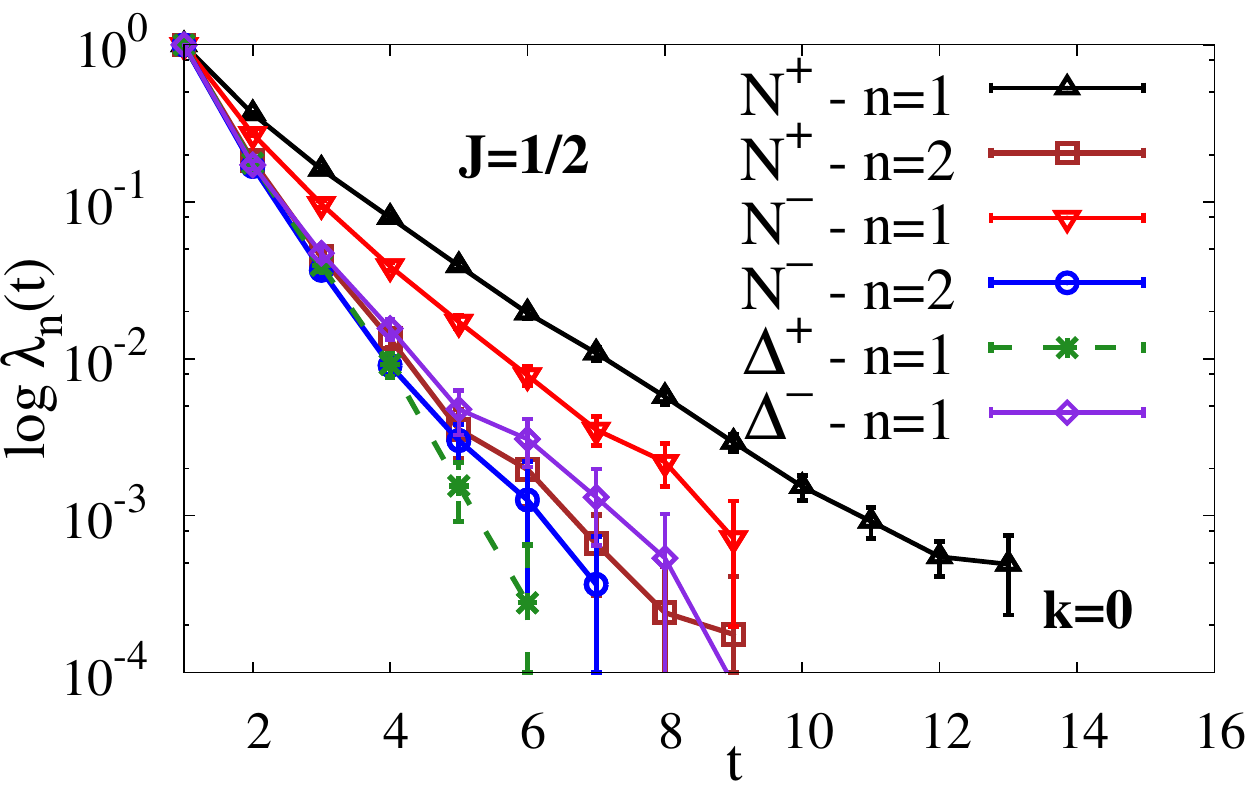}
        \caption{}
      \end{subfigure}
       \hspace*{-32pt}
      \begin{subfigure}[b]{0.45\textwidth}
          \centering
          \includegraphics[scale=0.5]{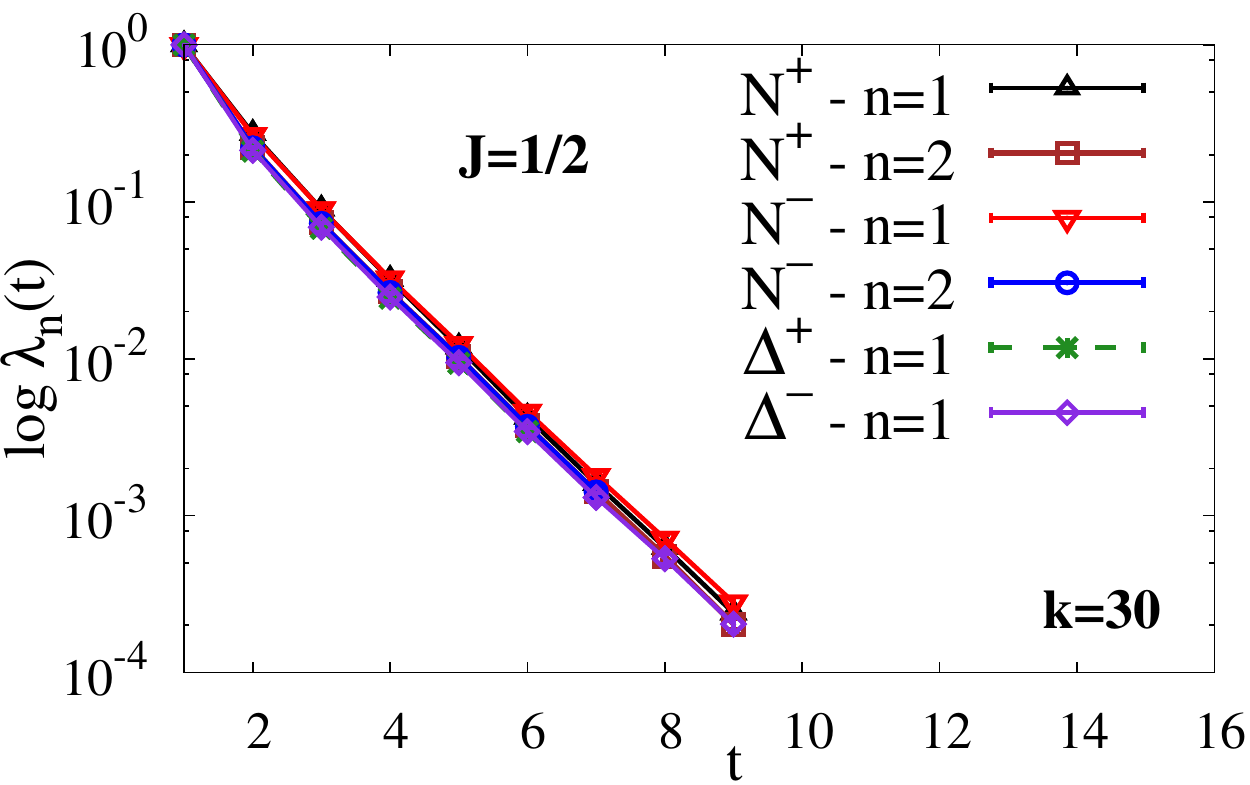}
          \caption{}
      \end{subfigure}
     \hspace*{24pt} \hfill\\
        \begin{subfigure}[b]{0.45\textwidth} 
           \includegraphics[scale=0.5]{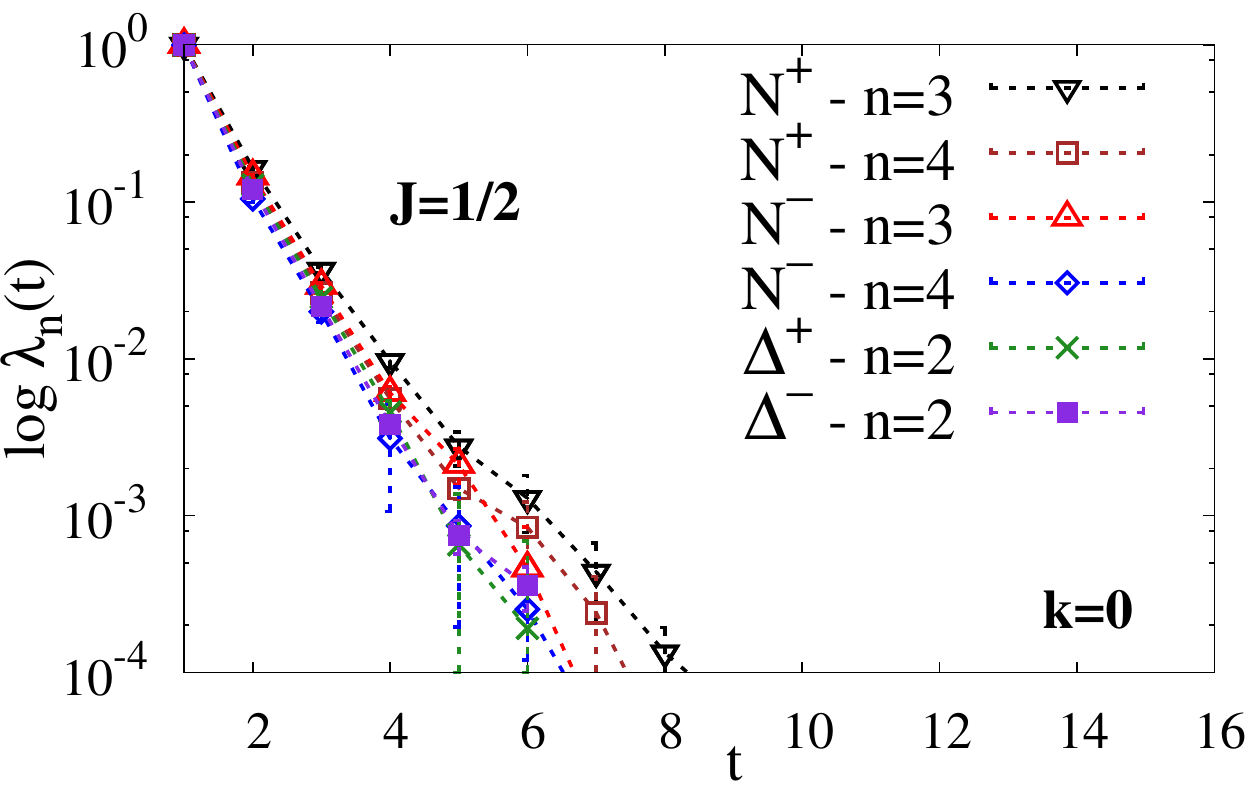}
        \caption{}
      \end{subfigure}
       \hspace*{-32pt}
      \begin{subfigure}[b]{0.45\textwidth}
          \centering
          \includegraphics[scale=0.5]{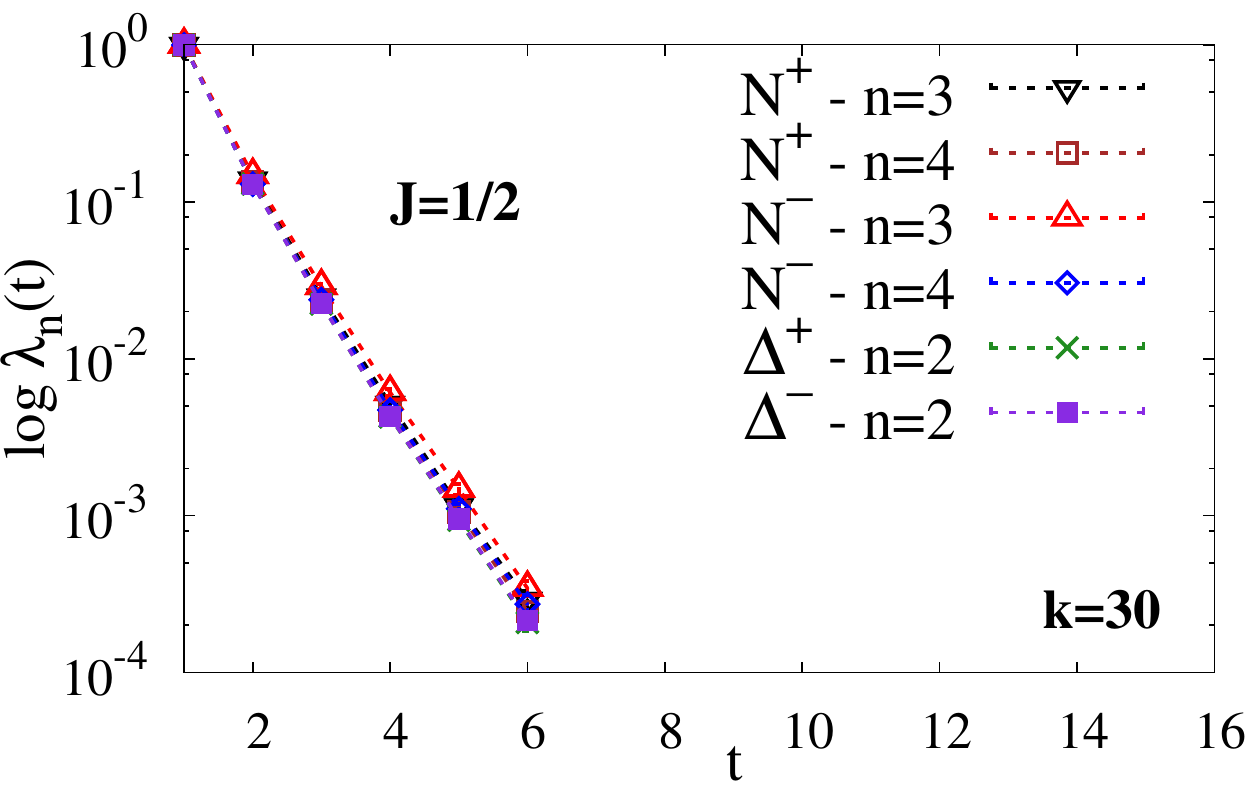}
          \caption{}
      \end{subfigure}
     \hspace*{24pt} \hfill\\  
      \caption{Eigenvalues of $J=\frac{1}{2}^{\pm}$ baryons: (a) $N$(n=1,2) and $\Delta$(n=1), 
      (c) $N$(n=3,4) and $\Delta$(n=2) in the full case ($k=0$);
     (b) $N$(n=1,2) and $\Delta$(n=1)  (d) $N$(n=3,4) and $\Delta$(n=2) after excluding $k=30$ 
quasi-zero modes.
}\label{fig:BB12ev}
   \end{figure*}
   
Now comes the crucial step.
From now on we address the chiralspin $SU(2)_{CS}$ and $SU(4)$ transformation
properties of the fields above. For definitions of these groups and respective
transformations we refer to \cite{Glozman:2015qva}.
The interpolators (\ref{Interpolator1}) and (\ref{Interpolator2}) cannot be connected via $SU(2)_A$ or $U(1)_A$ transformations.  However, the 
 $SU(2)_{CS}$ chiralspin rotation does connect respective diquarks,
 see Fig.~\ref{Table1}. Consequently, all nucleon fields $N(\frac{1}{2},{\frac{1}{2}}^{\pm})$ from Table \ref{tab:BI} are connected via $SU(2)_{CS}$.
 Via $SU(4)$ all diquarks  in Fig.~\ref{Table1} get
connected to each other. All nucleon fields $N(\frac{1}{2},{\frac{1}{2}}^{\pm})$ from Table \ref{tab:BI} are members of a 20-plet of $SU(4)$.

Given this result we can formulate the following prediction.
If correlation functions of both parities obtained with all three $N(\frac{1}{2},{\frac{1}{2}}^{\pm})$ interpolators
from Table II became identical (or with the interpolators 
(\ref{Interpolator1}) and (\ref{Interpolator2}) -
it is sufficient) and the respective states got degenerate, it 
would mean that all symmetries $SU(2)_L \times SU(2)_R$, $SU(2)_{CS} \supset U(1)_A$ and $SU(4)$ are restored.

The $\Delta(\frac{3}{2},\frac{1}{2}^\pm)$ interpolator from Table \ref{tab:BI} contains a $J=1$ diquark and is not connected with the $N(\frac{1}{2},\frac{1}{2}^\pm)$ interpolators from the
same Table via $U(1)_A$, $SU(2)_A$, $SU(2)_{CS}$ or $SU(4)$ transformations.
It belongs to a different 20-plet of $SU(4)$. A coincidence of its correlator
with the correlators of  $N(\frac{1}{2},\frac{1}{2}^\pm)$ and a degeneracy
of respective states would indicate a symmetry that is higher than $SU(4)$
and that contains $SU(4)$ as a subgroup.

   \begin{figure*}[htb]
    \centering 
    \begin{subfigure}[b]{0.45\textwidth}
        \centering
        \includegraphics[scale=0.5]{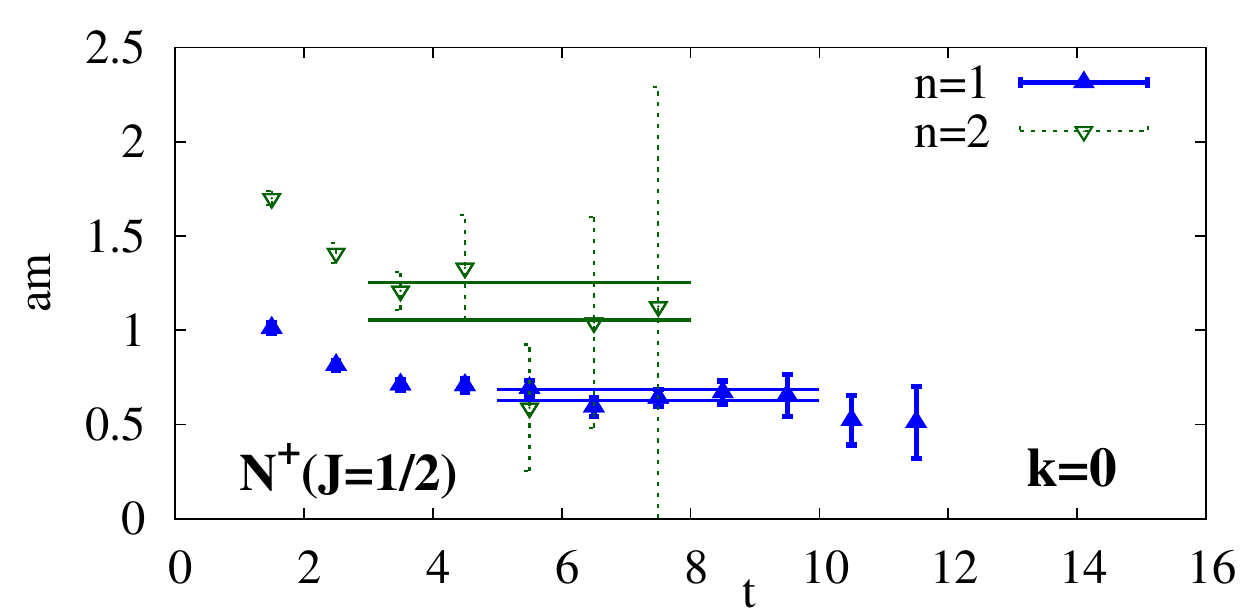}\\
         \includegraphics[scale=0.5]{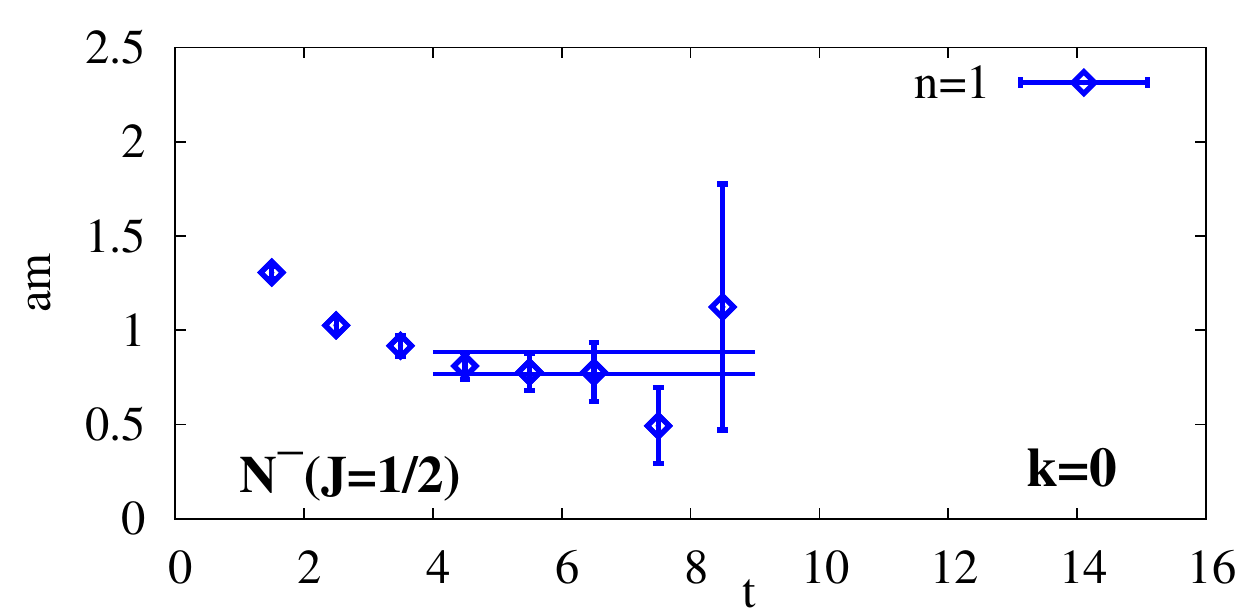}\\ 
         \includegraphics[scale=0.5]{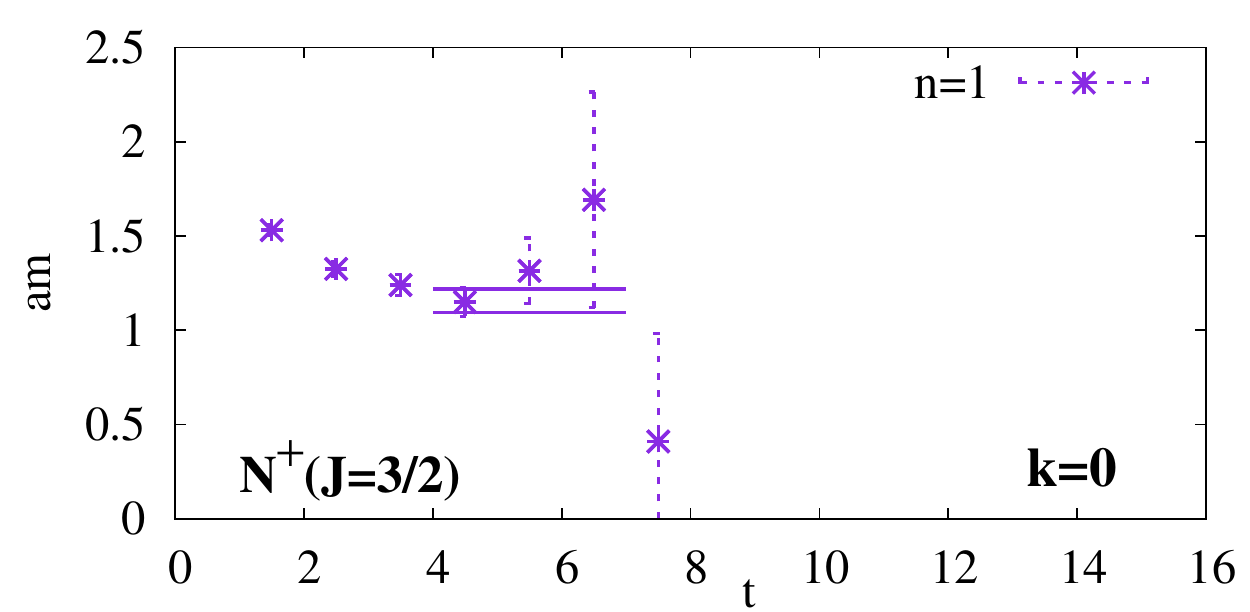}\\
         \includegraphics[scale=0.5]{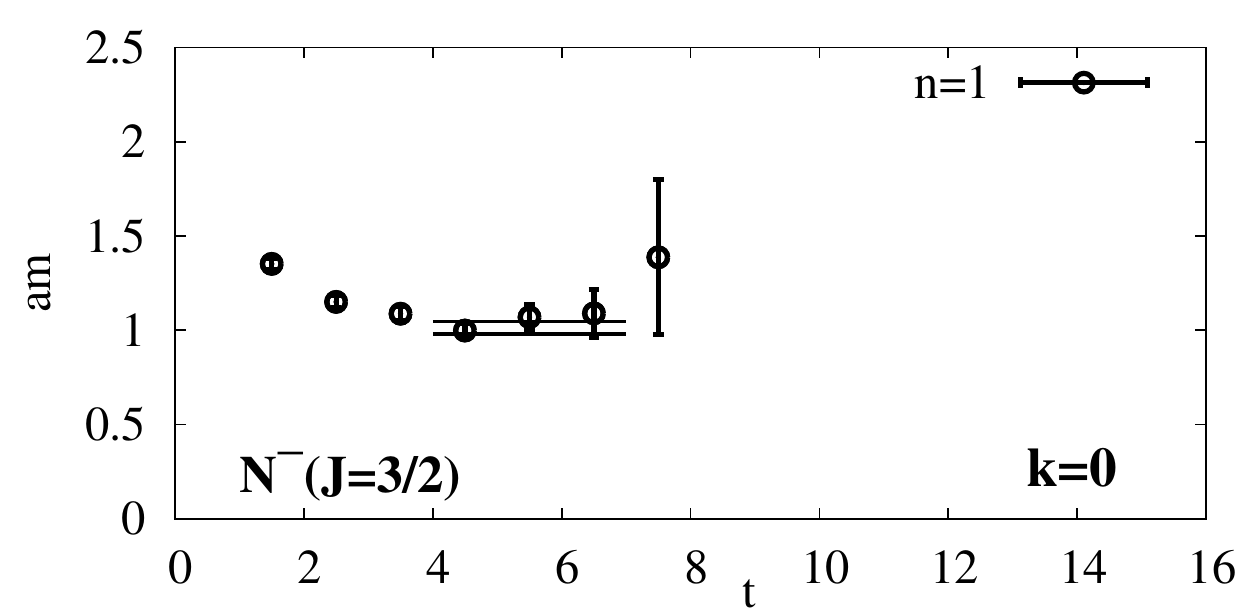}\\  
        
        \caption{}\label{fig:Bk0}
      \end{subfigure}
       \hspace*{-32pt}
      \begin{subfigure}[b]{0.45\textwidth}
          \centering
           \includegraphics[scale=0.5]{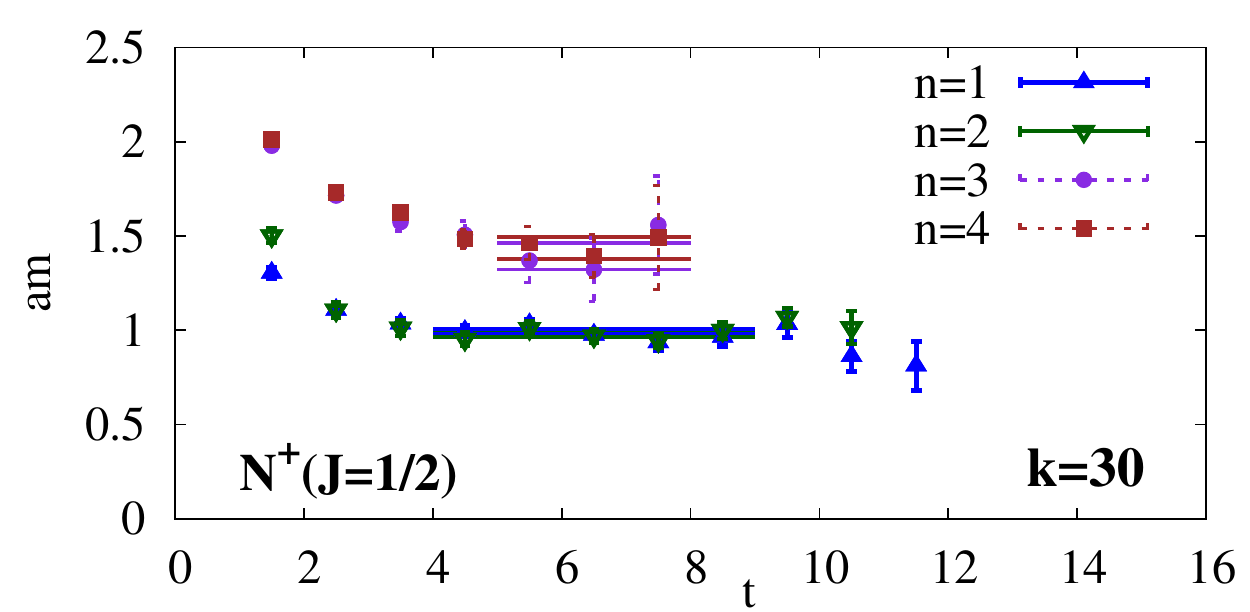}\\
         \includegraphics[scale=0.5]{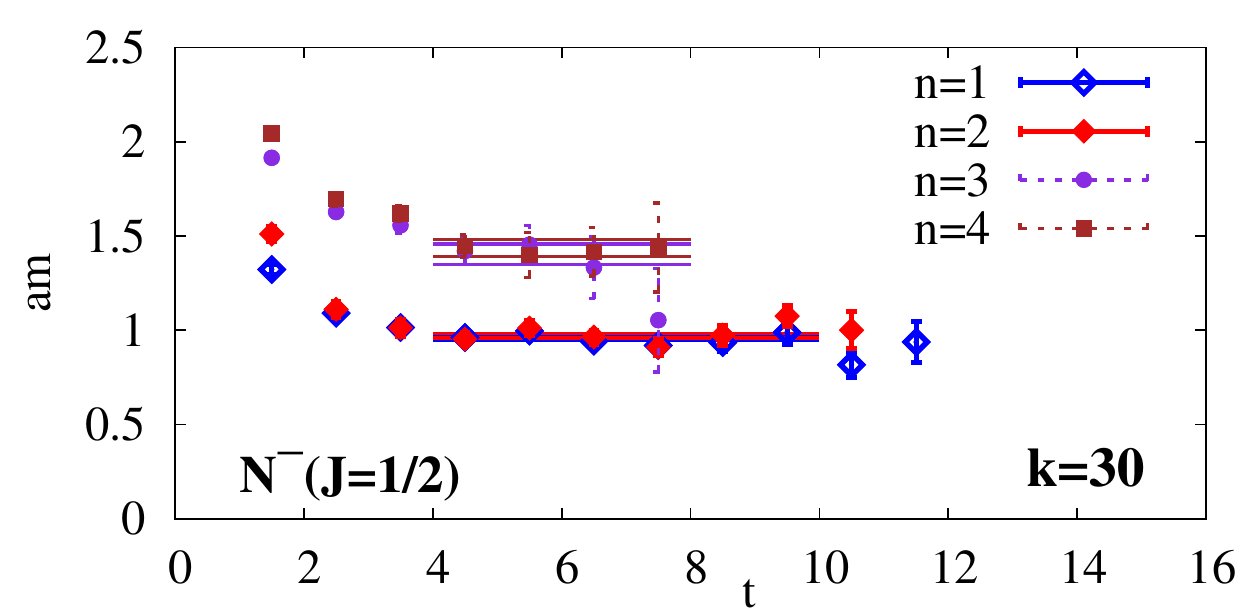}\\  
            \includegraphics[scale=0.5]{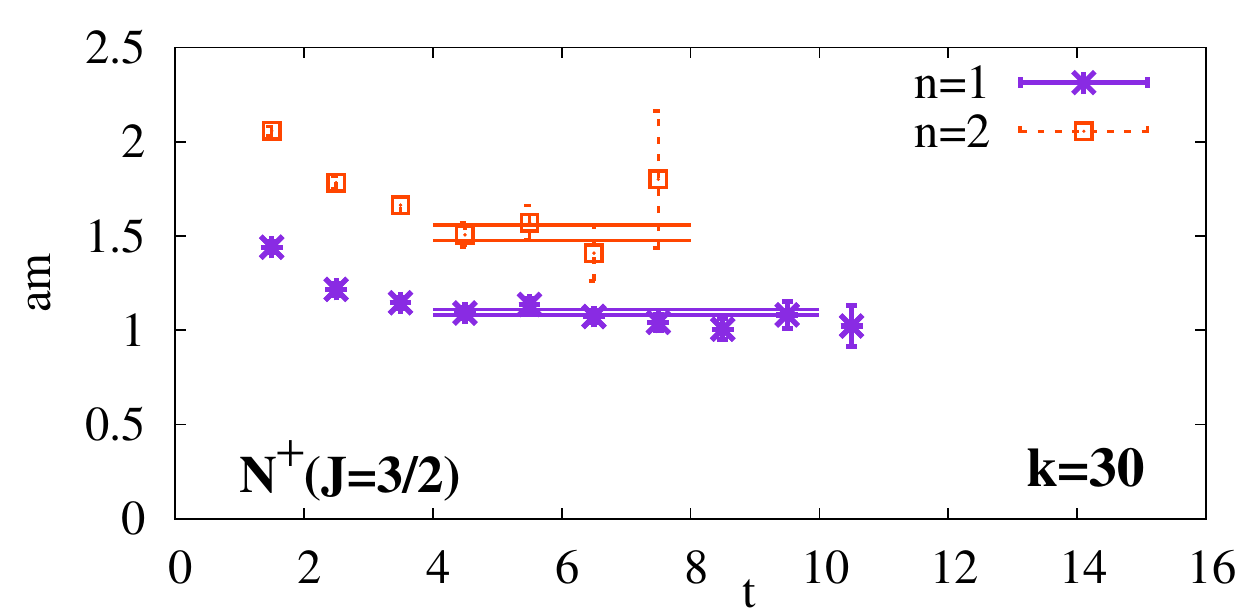}\\
         \includegraphics[scale=0.5]{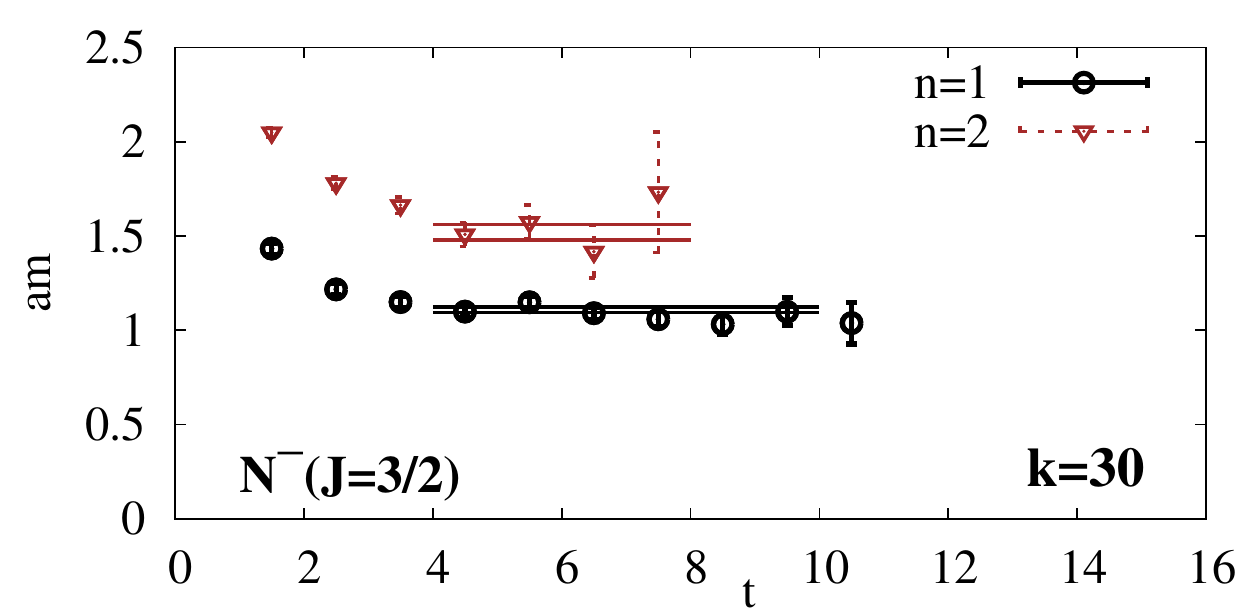}\\        
          \caption{}\label{fig:Bk30}
      \end{subfigure}
     \hspace*{24pt} \hfill\\
            \caption{ $N^\pm(J=\frac{1}{2})$,$N^\pm(J=\frac{3}{2})$: effective masses after excluding (a) $k=0$, (b)  $k=30$
quasi-zero modes.
}\label{fig:Nm}
   \end{figure*}
   
\section{Setup of the lattice simulation}
\label{Chapter-Lattice-Setup}
Our lattice is $16^3 \times 32$ with a lattice spacing $a\sim 0.12 $ fm. 
The dynamical $N_F = 2$ Overlap fermion gauge field configurations (in the topological sector 
$Q_{\textsc{T}} = 0$) were generously provided by the JLQCD collaboration, Refs.~\cite{Aoki:2008tq,Aoki:2012pma}. 
The pion mass is $M_{\pi} = 289(2)$ MeV, Ref.~\cite{Noaki:2008iy}. In our simulation
we use 87 gauge field configurations.

In our studies we use all operators from Table \ref{tab:BI} except the one in the second
line of this Table. A set of different extended sources with 
different smearing widths  allows for a larger operator
basis in the variational method.

The quasi-zero modes are removed from the inverse Overlap Dirac operator, denoted as $S_{\textsc{full}}(x,y)$, via 
\begin{align}
S_{k}(x,y) = S_{\textsc{full}}(x,y) -  \sum_{i=1}^k \frac{1}{\lambda_i} v_i(x)  v^{\dagger}_i(y) \, .
\end{align}
The low-mode truncated inverse Dirac operator $S_{k}(x,y)$ depends on the mode number $k$, 
which we have chosen at $k=2,4,10,16,20,30$.

The full (untruncated) Dirac operator is inverted using Jacobi smeared quark sources,  Ref.~\cite{Burch:2004he}, with two different 
smearing widths, which we refer to as "narrow" (n) and "wide" (w).  

The hadron states are extracted via the so-called variational
method, see Ref.~\cite{Michael:1985ne, Luscher:1990ck, Blossier:2009kd}.  One computes
the cross-correlation matrix $C_{ij}(t)= \langle \mathcal{O}_i(t) \mathcal{O}^{\dagger}_j(0)\rangle$,
whose entries are the interpolators (with different smearing widths), which generate a given state. 
From the solution to the generalized eigenvalue problem 
\begin{equation}\label{eq:GEVP}
 C(t) \vec{v}_n(t)= \lambda_n(t)C(t_0)\vec{v}_n(t) \; , 
\end{equation}
one obtains the eigenvalues $\lambda_n(t)$. An exponential decay 
\begin{equation}
\lambda^{(n)}(t,t_0) \propto \E^{-E_n (t-t_0)}(1+\mathcal{O}(\E^{-\Delta E_n (t-t_0)}),
\end{equation}
allows for the extraction of the mass of a state.

     \begin{figure*}[htb]
    \centering 
        \begin{subfigure}[b]{0.45\textwidth}
        \centering
         \includegraphics[scale=0.5]{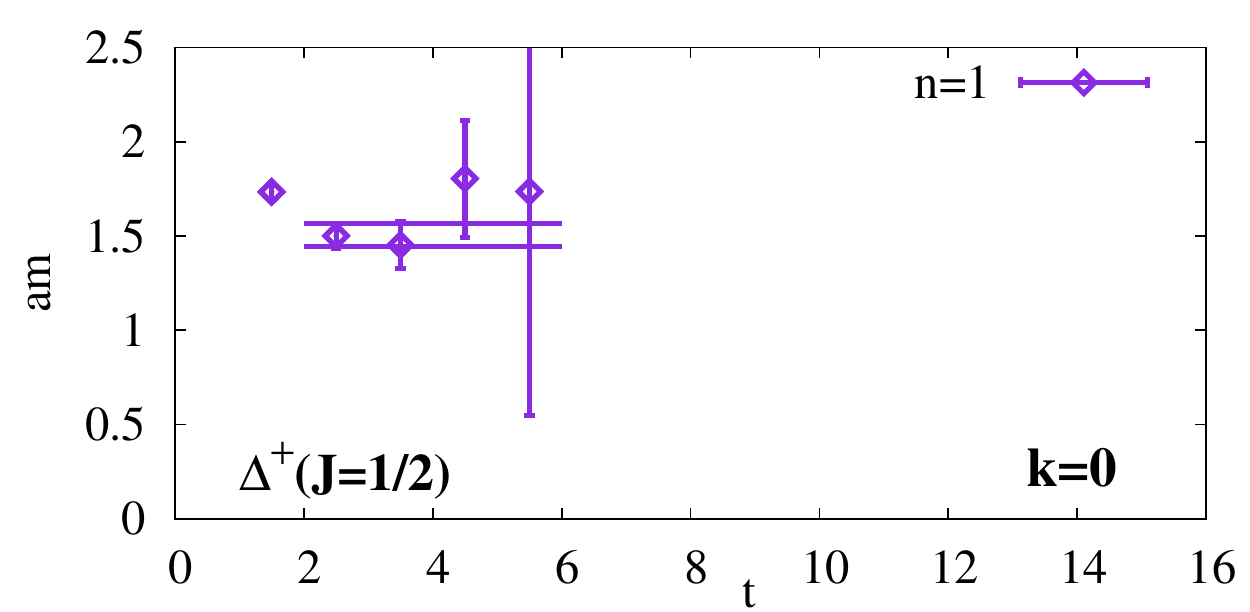}\\
         \includegraphics[scale=0.5]{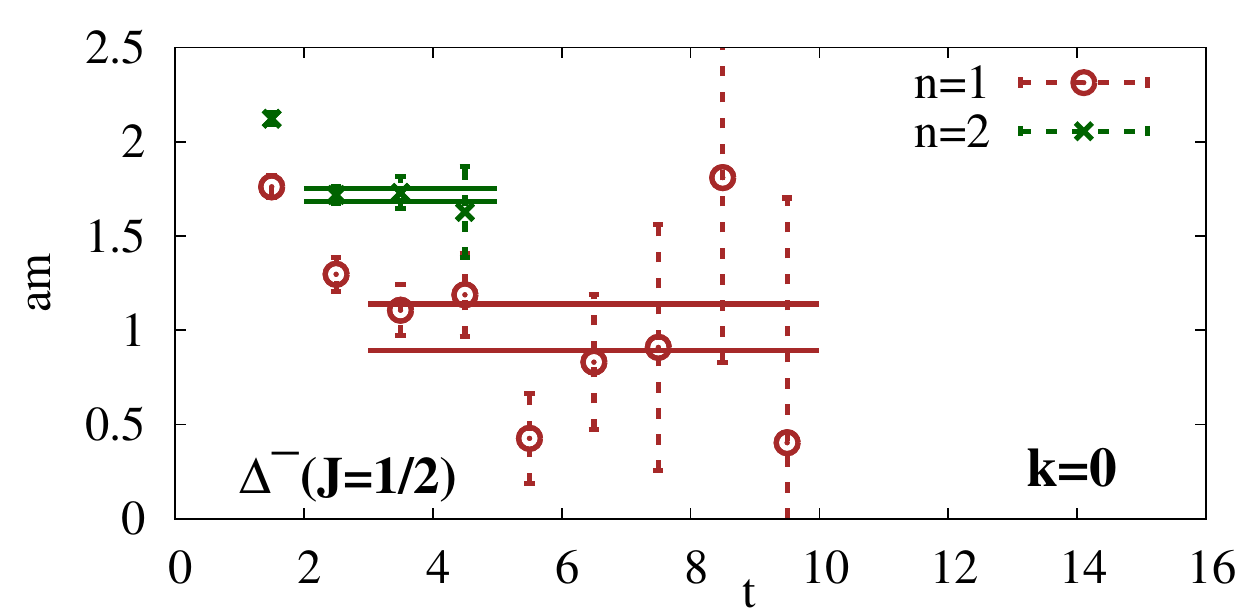}\\  
        \includegraphics[scale=0.5]{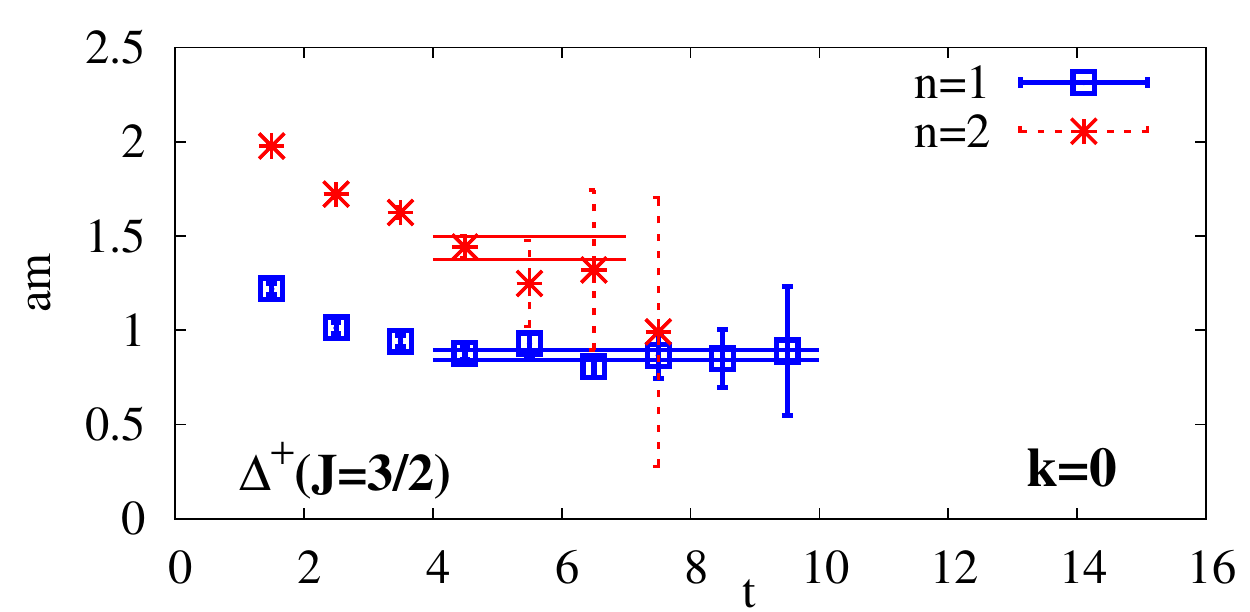}\\
         \includegraphics[scale=0.5]{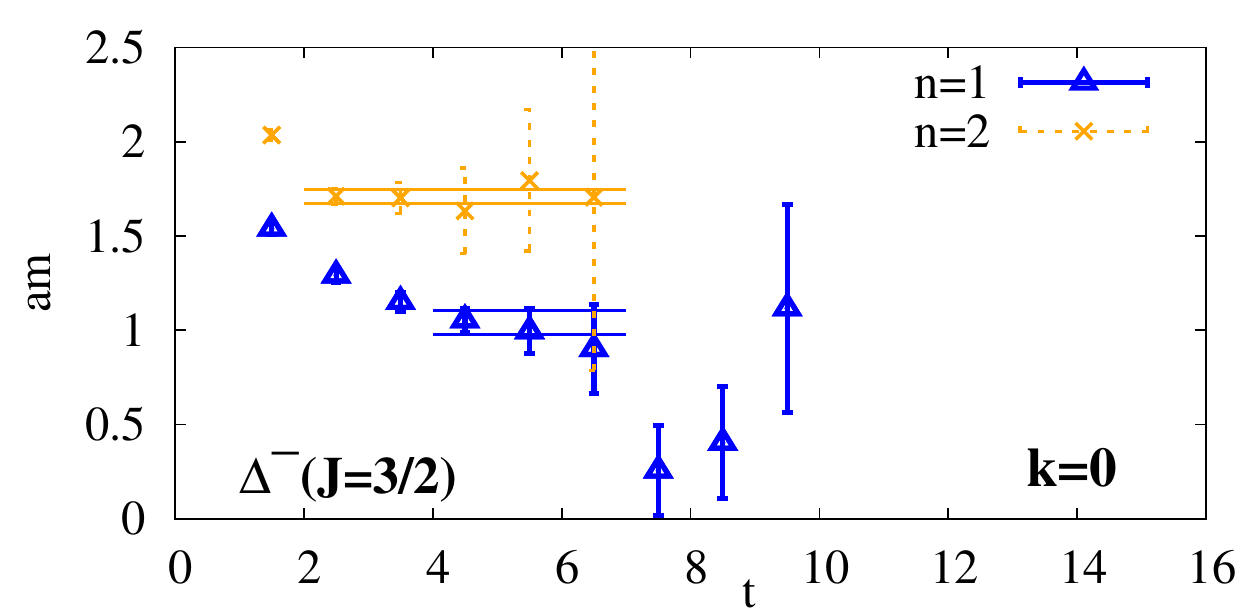}\\  
        \caption{}\label{fig:Dk0}
      \end{subfigure}
       \hspace*{-32pt}
      \begin{subfigure}[b]{0.45\textwidth}
          \centering
          \includegraphics[scale=0.5]{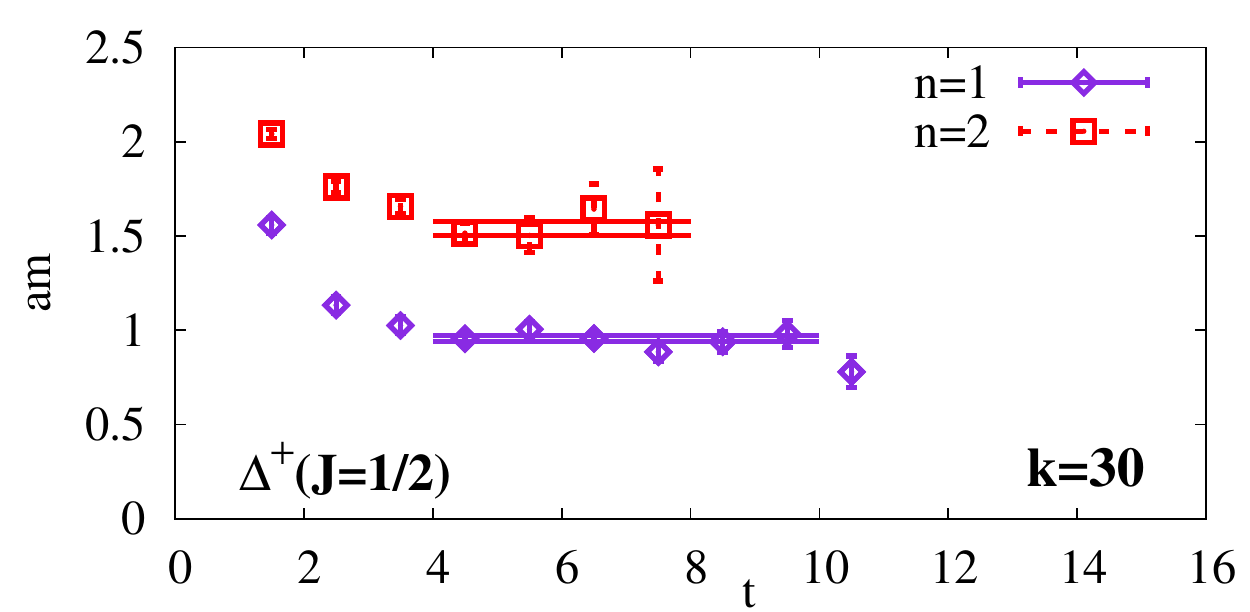}\\
         \includegraphics[scale=0.5]{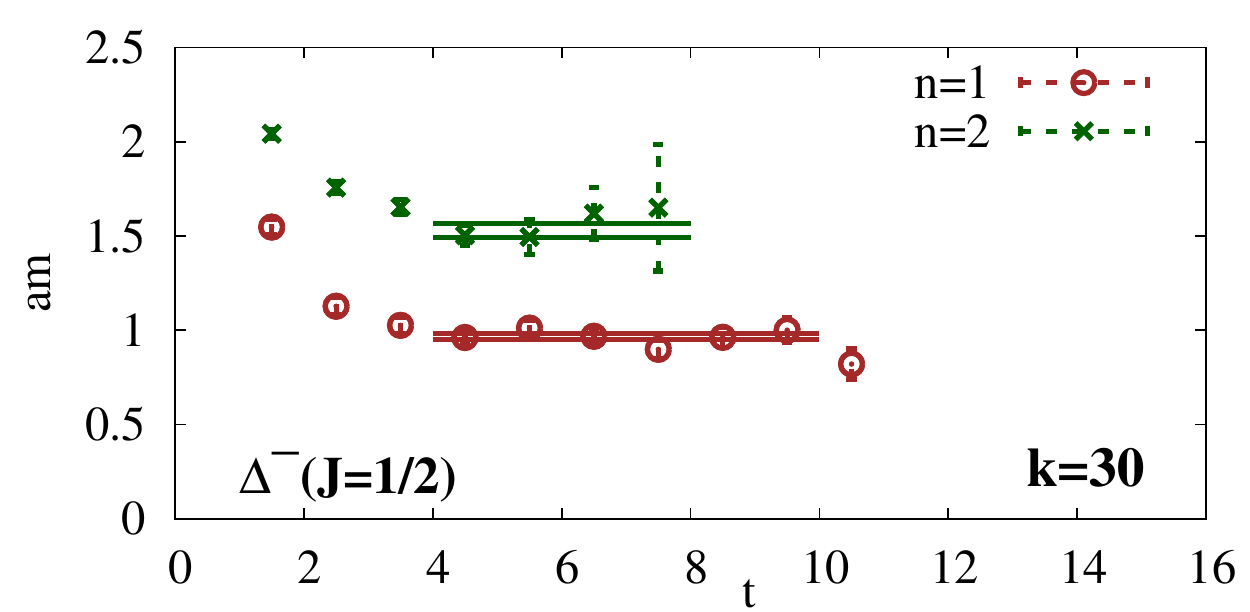}\\ 
           \includegraphics[scale=0.5]{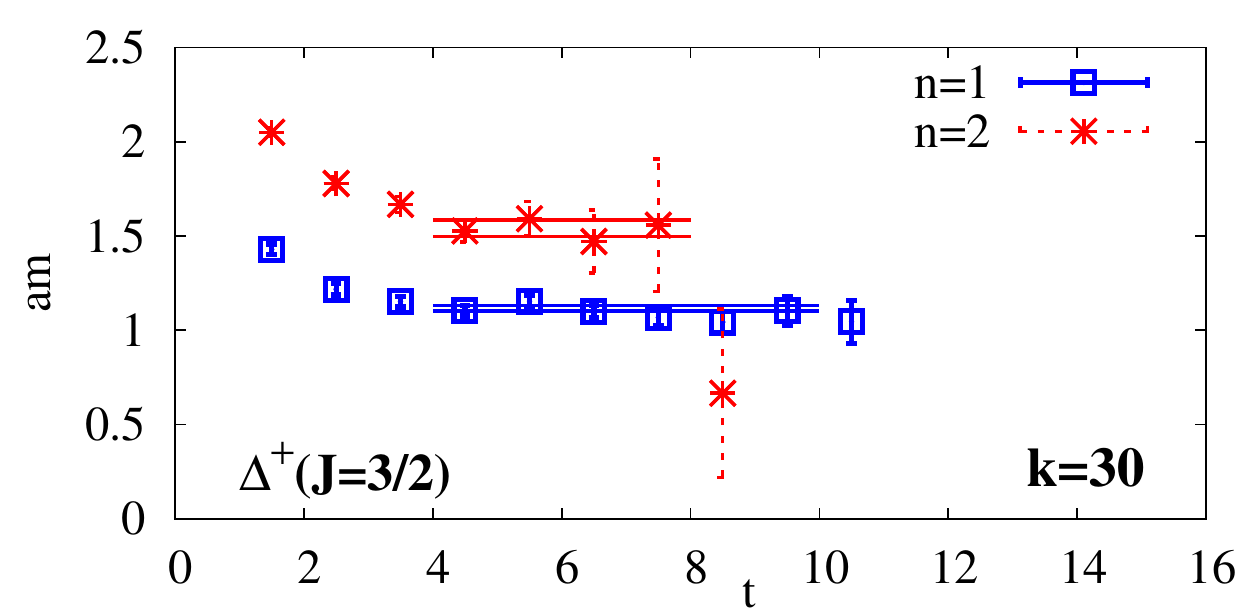}\\
         \includegraphics[scale=0.5]{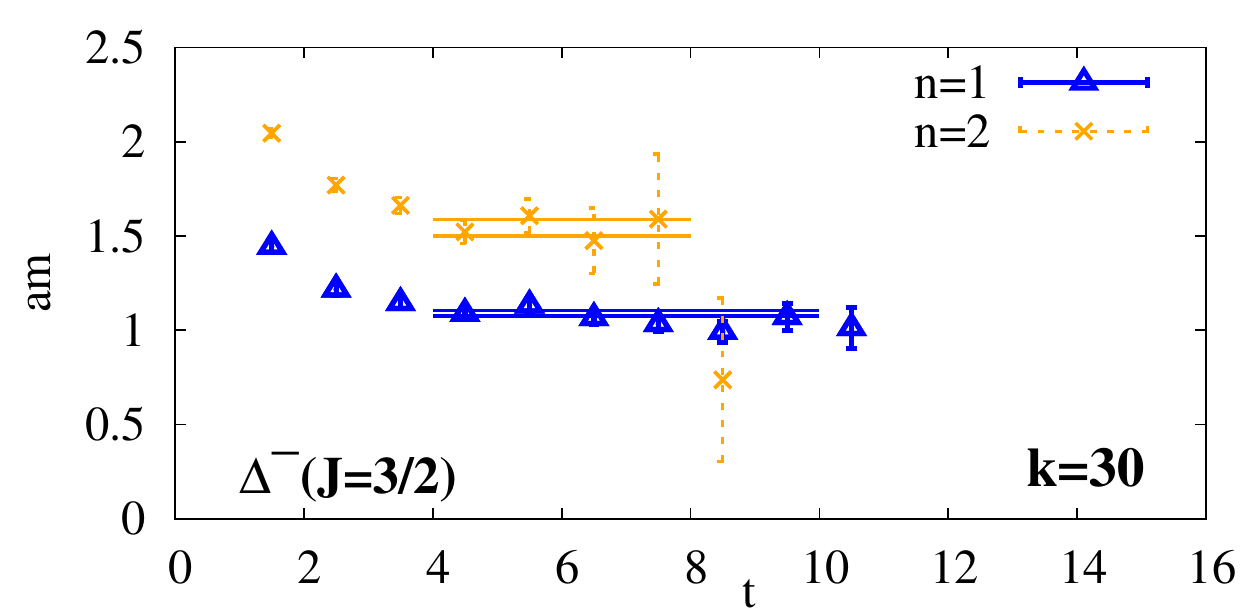}\\        
          \caption{}\label{fig:Dk30}
      \end{subfigure}
     \hspace*{24pt} \hfill\\
      
            \caption{ $\Delta^\pm(J=\frac{1}{2})$,$\Delta^\pm(J=\frac{3}{2})$: effective masses  after excluding (a) $k=0$, (b)  $k=30$
quasi-zero modes.
}\label{fig:Bm}
   \end{figure*}

         \begin{figure*}[t!]
    \centering 
          \includegraphics[scale=0.55]{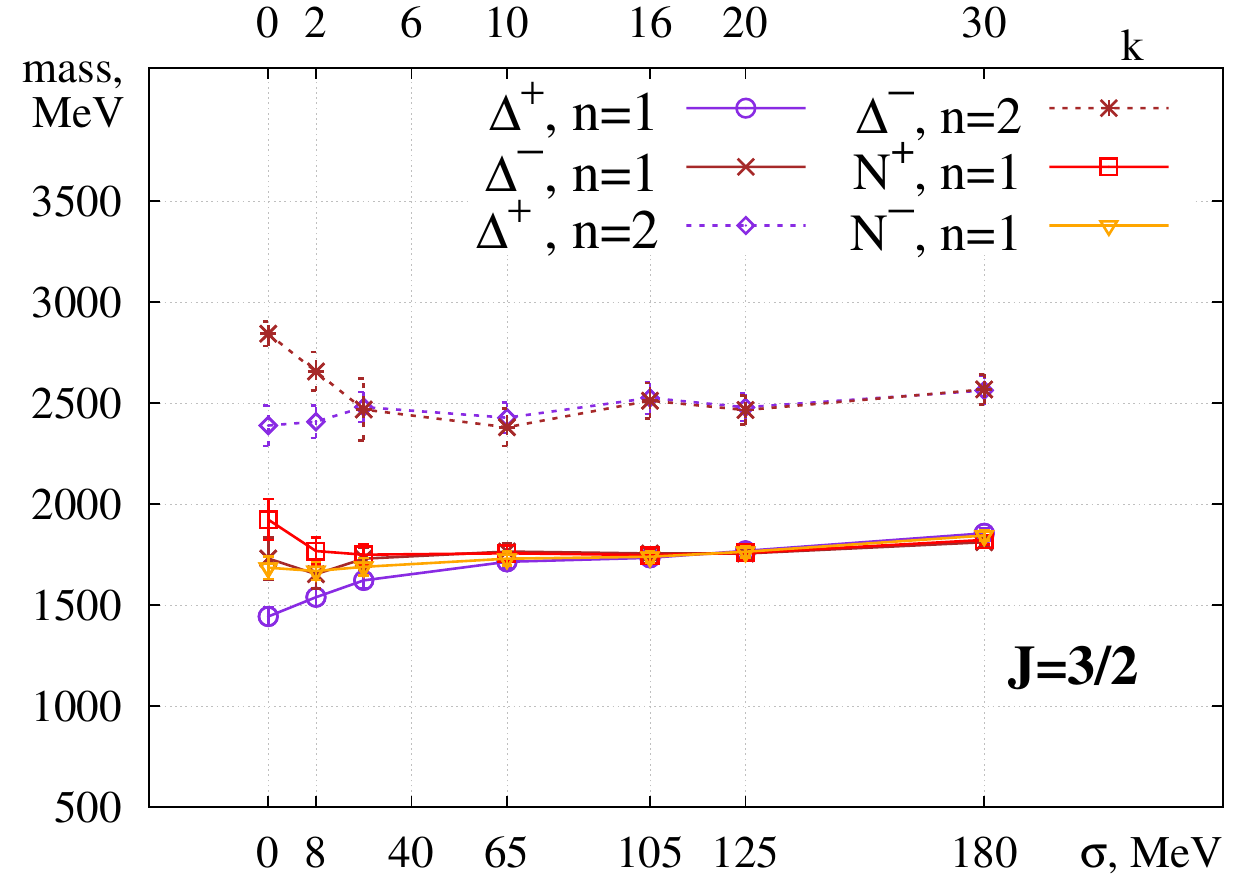}   
      \caption{Mass evolution of $N$ and $\Delta$ baryons with total spin $J=\frac{3}{2}$  on exclusion of the quasi-zero modes. The value
$\sigma$ denotes the energy gap.}\label{fig:BB32}
   \end{figure*}
   
         \begin{figure*}[t!]
    \centering 
        \includegraphics[scale=0.55]{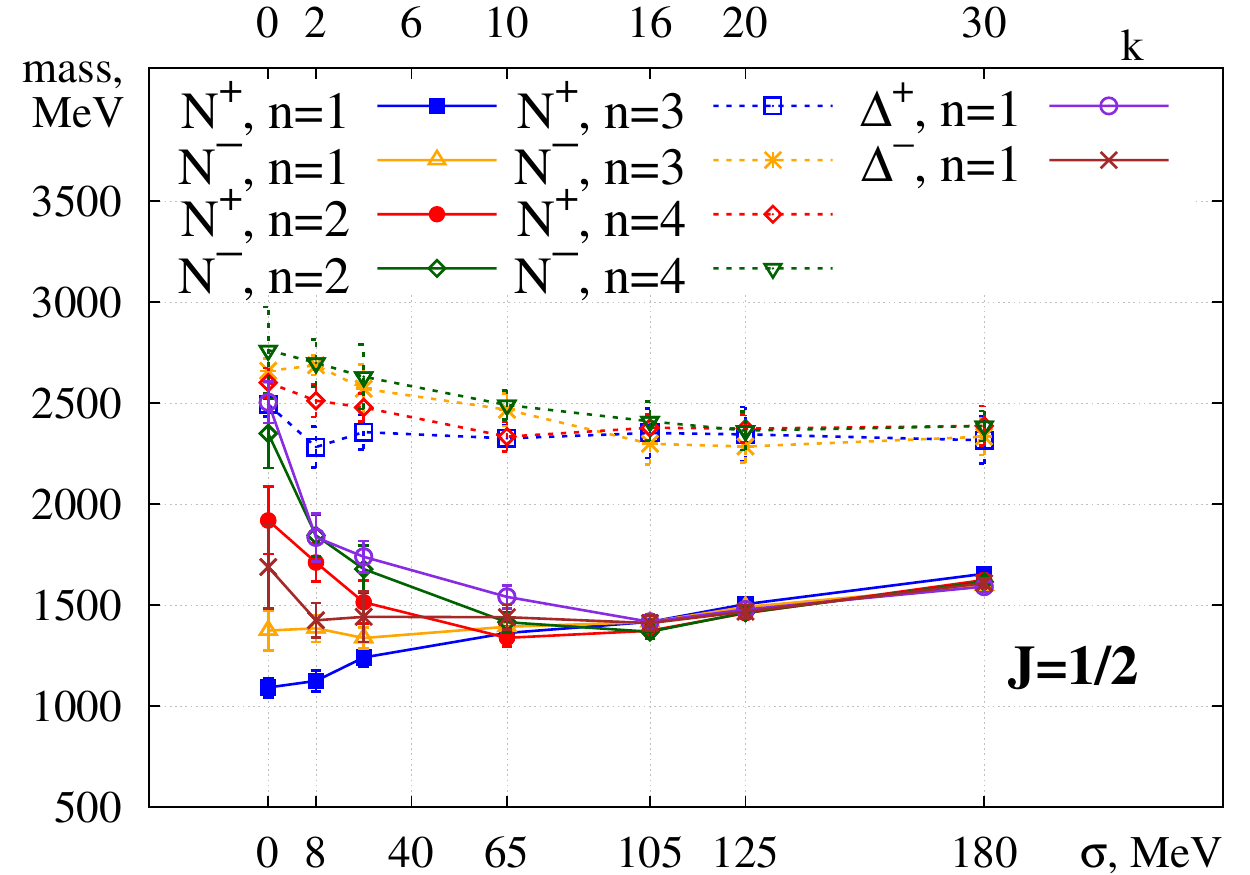}  
      \caption{Mass evolution of $N$ and $\Delta$ baryons with total spin  $J=\frac{1}{2}$ on exclusion of the quasi-zero modes. The value
$\sigma$ denotes the energy gap.}\label{fig:BB12}
   \end{figure*}

    \begin{figure*}[t!]
    \centering 
      \begin{subfigure}[b]{0.45\textwidth}
        \centering
        \includegraphics[scale=0.5]{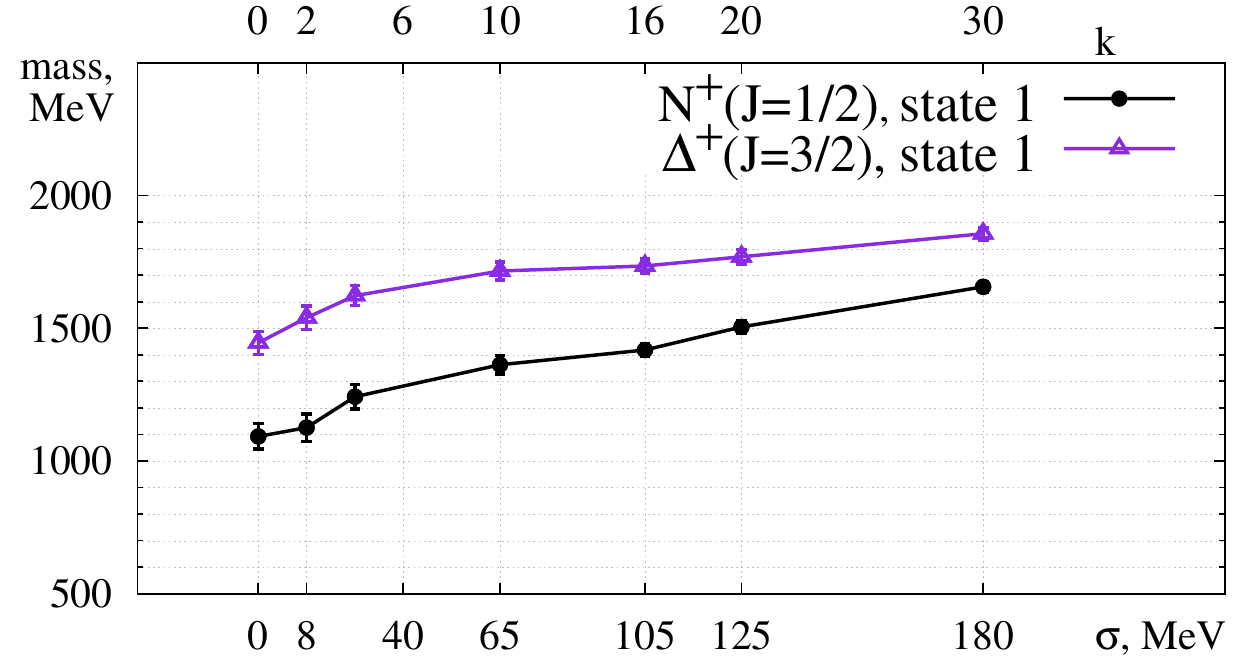}
        \caption{}
      \end{subfigure}
       \hspace*{-32pt}
      \begin{subfigure}[b]{0.45\textwidth}
          \centering
          \includegraphics[scale=0.5]{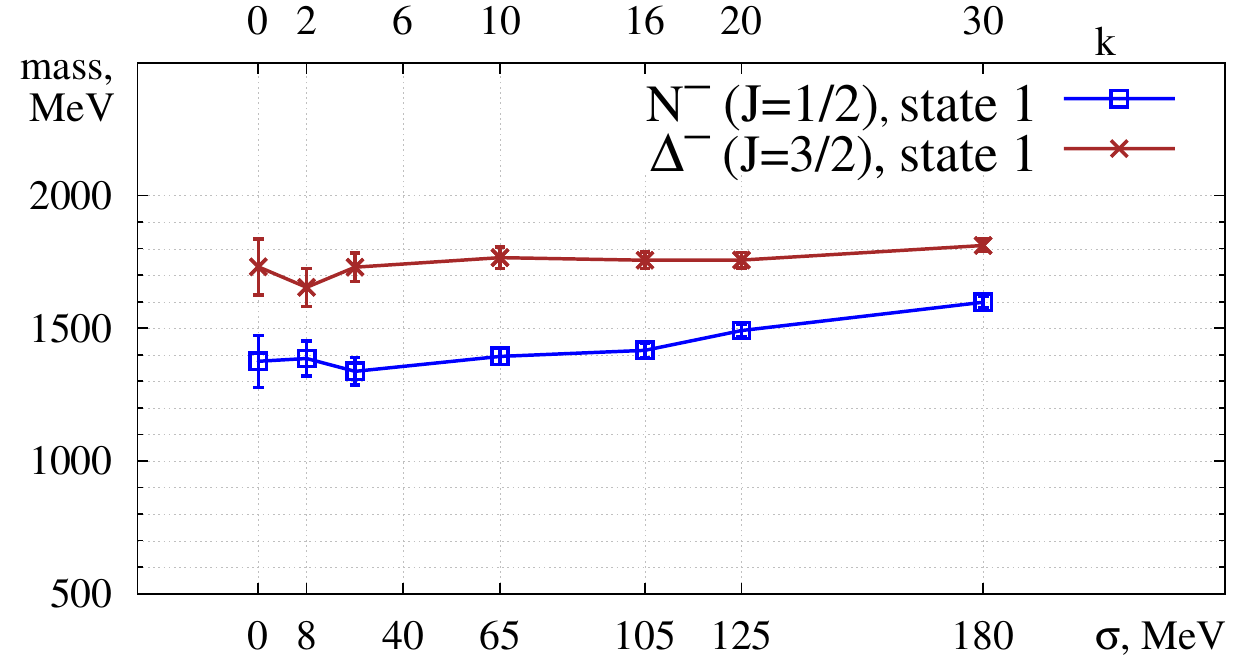}
          \caption{}
      \end{subfigure}
     \hspace*{24pt} \hfill\\ 
      \caption{Mass splittings of the $N(J=\frac{1}{2})$  and $\Delta(J=\frac{3}{2})$  baryons of (a) positive, (b) negative parities on exclusion of the low-lying modes.}\label{fig:BB2}
   \end{figure*}

\section{ Results}
\label{Chapter-Results}

In Fig.~\ref{fig:BB32ev} we show the eigenvalues of the correlation matrix for the $J=\frac{3}{2}^{\pm}$ case (the two last $J=3/2$ operators from Table \ref{tab:BI}). A coincidence of all four  correlators
signals restoration of the chiral  $SU(2)_L \times SU(2)_R$ symmetry.

In Fig.~\ref{fig:BB12ev} the eigenvalues of the correlation matrix for $J=\frac{1}{2}^{\pm}$ $N$ and $\Delta$ states are shown (the first, the third 
and the fourth operator from Table \ref{tab:BI}).
In the full, untruncated case they are different from each other. After removing the quasi-zero modes, \textit{all} eigenvalues fall on the same curve. This indicates the following 
 set of symmetries: $SU(2)_L \times SU(2)_R$, $SU(2)_{CS} \supset U(1)_A$ and $SU(4) \supset SU(2)_L \times SU(2)_R \times U(1)_A$.  It also indicates a larger symmetry that includes
 $SU(4)$ as a subgroup, as discussed at the end of Chapter \ref{Chapter-Chiral-Parity}.

In Fig.~\ref{fig:Nm} and Fig.~\ref{fig:Bm} we show effective mass
plots for all studied correlators and the fit ranges. 
Additional information on the fits, the mass values of the extracted states 
and the statistical errors can be found in Appendix \ref{Chapter-Appendix}.
 
What is common to all effective masses is that the signal improves after quasi-zero mode removal and the plateau becomes more stable. The
same behavior has also been observed for mesons.

In Figs.~\ref{fig:BB32} and \ref{fig:BB12} we show the evolution of
 masses of the $J=\frac{3}{2}$ and $J=\frac{1}{2}$ baryons 
 as a function of the
truncation parameter $k$ (or the energy gap $\sigma$). The onset of the degeneracy of states is observed between removal of $10-16$ modes like in
our previous studies of mesons. 

From the results presented in Fig.~\ref{fig:BB32} we can only conclude
about the $SU(2)_L \times SU(2)_R$ restoration, since the operators
$N(\frac{1}{2},{\frac{3}{2}}^{\pm})$ and
$\Delta(\frac{3}{2},{\frac{3}{2}}^{\pm})$  form
a $(1,\frac{1}{2})+(\frac{1}{2},1)$ - 12-plet of the parity-chiral group.

The degeneracy shown in Fig. ~\ref{fig:BB12} implies, however, a set
of some additional symmetries. A degeneracy of two positive- and two negative-parity
nucleons obtained with the operators (\ref{Interpolator1}) 
(first line of Table \ref{tab:BI}) and (\ref{Interpolator2}) (third line in Table II)
is a sufficient condition to claim a restoration of
$SU(2)_L \times SU(2)_R$, $SU(2)_{CS} \supset U(1)_A$ and $SU(4) \supset SU(2)_L \times SU(2)_R \times U(1)_A$ symmetries.

A larger degeneracy, seen in this figure, that involves also the $\Delta$
 states of both parities, generated with the fourth operator
 from  Table \ref{tab:BI}, implies some higher symmetry, that includes
 the $SU(4)$ group as a subgroup. The latter operator,
$\Delta(\frac{3}{2},\frac{1}{2}^\pm)$, is a member of a distinct dim=20
irreducible
representation of $SU(4)$ as compared to the interpolators
(\ref{Interpolator1}) and (\ref{Interpolator2}). The two different
irreducible dim=20 representations of $SU(4)$ can be connected
to each other only through some higher symmetry that includes $SU(4)$
as a subgroup. This situation is very similar to the $f_1$ (a singlet
of $SU(4)$) and $\rho, \rho', a_1, h_1, b_1, \omega, \omega'$ (a 15-plet
of $SU(4)$) degeneracy discussed at the end of Ref. \cite{Glozman:2015qva}.
We conclude that both meson and baryon data suggest a higher symmetry,
that will be discussed elsewhere \cite{CG}.

Finally, in Fig.~ \ref{fig:BB2} we show  ground states of both parities
of nucleon ($J=1/2$) and delta ($J=3/2$). It is clearly seen from this figure
that the nuclen-delta splitting persists after removal of the lowest-lying modes, though it becomes smaller. It implies the following: The $N - \Delta$
splitting observed in the truncated case cannot be due to the color-magnetic
interaction between the valence quarks, as suggested by the naive quark model,
because in this regime the $SU(4)$ symmetry is manifest and the quarks interact
with each other only through the color-electric field. Consequently a splitting
can be related only to a different rotational energy of the $J=1/2$ and $J=3/2$ states. A part of the $N - \Delta$ splitting in the untruncated world is
due to chiral symmetry breaking effects.

\section{Summary and Conclusions}
\label{Conclusions}
We have investigated the baryon spectrum with two degenerate flavors under quasi-zero mode removal. 
For $J=\frac{1}{2}$ baryons we have used interpolators in distinct chiral multiplets, which are therefore
not connected via a $SU(2)_L \times SU(2)_R$ transformation. From the fact that we see a high degeneracy of states, namely of  states connected by the
$SU(2)_{CS}$ and $SU(4)$ transformations,  we can conclude, that $SU(4)$ is also a symmetry of baryons after the quasi-zero mode removal. 

A degeneracy of baryons from different irreducible representations
of $SU(4)$ indicates the existence of a larger symmetry that contains $SU(4)$
as a subgroup.

\begin{acknowledgements}
We thank the JLQCD collaboration for supplying us with the Overlap
gauge configurations.
We acknowledge support from the Austrian Science Fund (FWF)
through the grants P26627-N27 and DK W1203-N16. The calculations have been performed on clusters 
at ZID at the University of Graz and 
at the Graz University of Technology.
\end{acknowledgements}

\begin{appendix}
 \section{\label{sec:ff} Masses and Fits}
 \label{Chapter-Appendix}
Masses of the baryon states are extracted by a single exponential fit to the eigenvalues. Corresponding mass values, fit ranges and statistics on 
 $\chi^2/$d.o.f.  analysis are presented in \tab{Tablek0} for truncations $k = 0, 16, 20, 30$.

  \begin{table*}[t!]
\begin{center}
\renewcommand{\arraystretch}{1.2}
\begin{tabular}{lccccclcccccc}
 \hline
 \hline
 \multicolumn{6}{c}{$ k = 0 $} & & \multicolumn{6}{c}{$\ k = 16 $} \\\cline{1-6}\cline{8-13} 
\multicolumn{1}{c}{   ($I$, $J^{PC}$)  } &\multicolumn{1}{c}{   $n$  } & $am$   &$\chi^2/d.o.f $ & $t$ & \#$i$ & &\multicolumn{1}{c}{   ($I$, $J^{PC}$)  } &\multicolumn{1}{c}{   $n$  } & $am$   &
$\chi^2/d.o.f $ & $t$ & \#$i$ \\\hline
\multicolumn{1}{l}{\multirow{2}{*}{$N(1/2,1/2^+)$}} & 1 & 0.657  $\pm$  0.029  & 3.11/4  & 5 - 10 & \multicolumn{1}{c}{\multirow{2}{*}{6}} & & \multicolumn{1}{l}{\multirow{2}{*}{$N(1/2,1/2^+)$}} & 1
& 0.852  $\pm$  0.015  & 2.21/4  & 6 - 11 & \multicolumn{1}{c}{\multirow{2}{*}{6}} \\
\multicolumn{1}{c}{ } & 2  & 1.154  $\pm$  0.100  & 3.67/4  & 3 - 8 & & &\multicolumn{1}{c}{ } & 2  & 0.826  $\pm$  0.020  & 9.73/5  & 5 - 11 &\\
\multicolumn{1}{l}{\multirow{2}{*}{$N(1/2,1/2^-)$}} & 1 & 0.827  $\pm$  0.059  & 2.11/4  & 4 - 9 & \multicolumn{1}{c}{\multirow{2}{*}{6}} & & \multicolumn{1}{l}{\multirow{2}{*}{$N(1/2,1/2^-)$}} & 1 &
0.851  $\pm$  0.016  & 7.14/5  & 5 - 11 & \multicolumn{1}{c}{\multirow{2}{*}{6}} \\
\multicolumn{1}{c}{ } & 2  & 1.412  $\pm$  0.102  & 1.80/3  & 3 - 7 & & &\multicolumn{1}{c}{ } & 2  & 0.823  $\pm$  0.020  & 6.75/5  & 5 - 11 &\\
\multicolumn{1}{l}{\multirow{1}{*}{$\Delta (3/2,1/2^+)$}} & 1 & 1.506  $\pm$  0.061  & 1.77/3  & 2 - 6 & \multicolumn{1}{c}{\multirow{1}{*}{3}} & & \multicolumn{1}{l}{\multirow{1}{*}{$\Delta
(3/2,1/2^+)$}} & 1 & 0.852  $\pm$  0.023  & 16.37/6  & 4 - 11 & \multicolumn{1}{c}{\multirow{1}{*}{3}} \\
\multicolumn{1}{l}{\multirow{1}{*}{$\Delta (3/2,1/2^-)$}} & 1 & 1.015  $\pm$  0.124  & 6.95/6  & 3 - 10 & \multicolumn{1}{c}{\multirow{1}{*}{3}} & & \multicolumn{1}{l}{\multirow{1}{*}{$\Delta
(3/2,1/2^-)$}} & 1 & 0.849  $\pm$  0.023  & 14.00/6  & 4 - 11 & \multicolumn{1}{c}{\multirow{1}{*}{3}} \\
\multicolumn{1}{l}{\multirow{1}{*}{$N(1/2,3/2^+)$}} & 1 & 1.157  $\pm$  0.061  & 3.45/2  & 4 - 7 & \multicolumn{1}{c}{\multirow{1}{*}{3}} & & \multicolumn{1}{l}{\multirow{1}{*}{$N(1/2,3/2^+)$}} & 1 &
1.051  $\pm$  0.017  & 10.03/6  & 4 - 11 & \multicolumn{1}{c}{\multirow{1}{*}{3}} \\
\multicolumn{1}{l}{\multirow{1}{*}{$N(1/2,3/2^-)$}} & 1 & 1.013  $\pm$  0.034  & 1.49/2  & 4 - 7 & \multicolumn{1}{c}{\multirow{1}{*}{3}} & & \multicolumn{1}{l}{\multirow{1}{*}{$N(1/2,3/2^-)$}} & 1 &
1.046  $\pm$  0.016  & 10.39/6  & 4 - 11 & \multicolumn{1}{c}{\multirow{1}{*}{3}} \\
\multicolumn{1}{l}{\multirow{2}{*}{$\Delta (3/2,3/2^-)$}} & 1 & 0.868  $\pm$  0.026  & 2.00/5  & 4 - 10 & \multicolumn{1}{c}{\multirow{2}{*}{3}} & & \multicolumn{1}{l}{\multirow{2}{*}{$\Delta
(3/2,3/2^-)$}} & 1 & 1.043  $\pm$  0.016  & 10.30/6  & 4 - 11 & \multicolumn{1}{c}{\multirow{2}{*}{3}} \\
\multicolumn{1}{c}{ } & 2  & 1.436  $\pm$  0.060  & 0.74/2  & 4 - 7 & & &\multicolumn{1}{c}{ } & 2  & 1.518  $\pm$  0.047  & 0.15/3  & 4 - 8 &\\
\multicolumn{1}{l}{\multirow{2}{*}{$\Delta (3/2,3/2^-)$}} & 1 & 1.040  $\pm$  0.063  & 0.38/2  & 4 - 7 & \multicolumn{1}{c}{\multirow{2}{*}{3}} & & \multicolumn{1}{l}{\multirow{2}{*}{$\Delta
(3/2,3/2^-)$}} & 1 & 1.055  $\pm$  0.018  & 8.68/4  & 4 - 9 & \multicolumn{1}{c}{\multirow{2}{*}{3}} \\
\multicolumn{1}{c}{ } & 2  & 1.710  $\pm$  0.036  & 0.13/4  & 2 - 7 & & &\multicolumn{1}{c}{ } & 2  & 1.510  $\pm$  0.052  & 0.22/3  & 4 - 8 &\\

\hline
 \hline
 \multicolumn{6}{c}{$ k = 20 $} & & \multicolumn{6}{c}{$\ k = 30 $} \\\cline{1-6}\cline{8-13} 
\multicolumn{1}{c}{  ($I$, $J^{PC}$) } &\multicolumn{1}{c}{   $n$  } & $am$   &$\chi^2/d.o.f $ & $t$ & \#$i$ & &\multicolumn{1}{c}{  ($I$, $J^{PC}$) } &\multicolumn{1}{c}{   $n$  } & $am$   &
$\chi^2/d.o.f $ & $t$ & \#$i$ \\\hline
\multicolumn{1}{l}{\multirow{4}{*}{$N(1/2,1/2^+)$}} & 1 & 0.995  $\pm$  0.012  & 4.81/4  & 4 - 9 & \multicolumn{1}{c}{\multirow{4}{*}{6}} & & \multicolumn{1}{l}{\multirow{4}{*}{$N(1/2,1/2^+)$}} & 1 &
0.905  $\pm$  0.014  & 7.26/7  & 5 - 13 & \multicolumn{1}{c}{\multirow{4}{*}{6}} \\
\multicolumn{1}{c}{ } & 2  & 0.976  $\pm$  0.013  & 1.73/4  & 4 - 9 & & &\multicolumn{1}{c}{ } & 2  & 0.878  $\pm$  0.015  & 11.48/7  & 5 - 13 &\\
\multicolumn{1}{c}{ } & 3  & 1.392  $\pm$  0.070  & 0.44/2  & 5 - 8 & & &\multicolumn{1}{c}{ } & 3  & 1.410  $\pm$  0.081  & 7.39/4  & 4 - 9 &\\
\multicolumn{1}{c}{ } & 4  & 1.435  $\pm$  0.059  & 0.18/2  & 5 - 8 & & &\multicolumn{1}{c}{ } & 4  & 1.427  $\pm$  0.041  & 5.59/4  & 4 - 9 &\\
\multicolumn{1}{l}{\multirow{4}{*}{$N(1/2,1/2^-)$}} & 1 & 0.960  $\pm$  0.013  & 4.68/5  & 4 - 10 & \multicolumn{1}{c}{\multirow{4}{*}{6}} & & \multicolumn{1}{l}{\multirow{4}{*}{$N(1/2,1/2^-)$}} & 1
& 0.896  $\pm$  0.013  & 6.50/6  & 4 - 11 & \multicolumn{1}{c}{\multirow{4}{*}{6}} \\
\multicolumn{1}{c}{ } & 2  & 0.971  $\pm$  0.013  & 9.83/5  & 4 - 10 & & &\multicolumn{1}{c}{ } & 2  & 0.879  $\pm$  0.017  & 10.67/6  & 4 - 11 &\\
\multicolumn{1}{c}{ } & 3  & 1.403  $\pm$  0.055  & 2.36/3  & 4 - 8 & & &\multicolumn{1}{c}{ } & 3  & 1.374  $\pm$  0.048  & 2.48/3  & 4 - 8 &\\
\multicolumn{1}{c}{ } & 4  & 1.435  $\pm$  0.045  & 0.23/3  & 4 - 8 & & &\multicolumn{1}{c}{ } & 4  & 1.421  $\pm$  0.057  & 2.52/4  & 4 - 9 &\\
\multicolumn{1}{l}{\multirow{2}{*}{$\Delta (3/2,1/2^+)$}} & 1 & 0.957  $\pm$  0.016  & 3.69/5  & 4 - 10 & \multicolumn{1}{c}{\multirow{2}{*}{3}} & & \multicolumn{1}{l}{\multirow{2}{*}{$\Delta
(3/2,1/2^+)$}} & 1 & 0.889  $\pm$  0.023  & 6.96/4  & 4 - 9 & \multicolumn{1}{c}{\multirow{2}{*}{3}} \\
\multicolumn{1}{c}{ } & 2  & 1.540  $\pm$  0.036  & 0.77/3  & 4 - 8 & & &\multicolumn{1}{c}{ } & 2  & 1.437  $\pm$  0.043  & 2.30/3  & 4 - 8 &\\
\multicolumn{1}{l}{\multirow{2}{*}{$\Delta (3/2,1/2^-)$}} & 1 & 0.967  $\pm$  0.015  & 3.50/5  & 4 - 10 & \multicolumn{1}{c}{\multirow{2}{*}{3}} & & \multicolumn{1}{l}{\multirow{2}{*}{$\Delta
(3/2,1/2^-)$}} & 1 & 0.882  $\pm$  0.020  & 6.78/4  & 4 - 9 & \multicolumn{1}{c}{\multirow{2}{*}{3}} \\
\multicolumn{1}{c}{ } & 2  & 1.528  $\pm$  0.038  & 0.92/3  & 4 - 8 & & &\multicolumn{1}{c}{ } & 2  & 1.429  $\pm$  0.046  & 2.24/3  & 4 - 8 &\\
\multicolumn{1}{l}{\multirow{2}{*}{$N(1/2,3/2^+)$}} & 1 & 1.096  $\pm$  0.014  & 7.90/5  & 4 - 10 & \multicolumn{1}{c}{\multirow{2}{*}{3}} & & \multicolumn{1}{l}{\multirow{2}{*}{$N(1/2,3/2^+)$}} & 1
& 1.057  $\pm$  0.018  & 9.67/4  & 4 - 9 & \multicolumn{1}{c}{\multirow{2}{*}{3}} \\
\multicolumn{1}{c}{ } & 2  & 1.518  $\pm$  0.042  & 1.22/3  & 4 - 8 & & &\multicolumn{1}{c}{ } & 2  & 1.481  $\pm$  0.044  & 1.67/3  & 4 - 8 &\\
\multicolumn{1}{l}{\multirow{2}{*}{$N(1/2,3/2^-)$}} & 1 & 1.109  $\pm$  0.014  & 6.37/5  & 4 - 10 & \multicolumn{1}{c}{\multirow{2}{*}{3}} & & \multicolumn{1}{l}{\multirow{2}{*}{$N(1/2,3/2^-)$}} & 1
& 1.061  $\pm$  0.017  & 8.74/4  & 4 - 9 & \multicolumn{1}{c}{\multirow{2}{*}{3}} \\
\multicolumn{1}{c}{ } & 2  & 1.520  $\pm$  0.040  & 1.05/3  & 4 - 8 & & &\multicolumn{1}{c}{ } & 2  & 1.471  $\pm$  0.042  & 1.50/3  & 4 - 8 &\\
\multicolumn{1}{l}{\multirow{2}{*}{$\Delta (3/2,3/2^-)$}} & 1 & 1.116  $\pm$  0.014  & 5.69/5  & 4 - 10 & \multicolumn{1}{c}{\multirow{2}{*}{3}} & & \multicolumn{1}{l}{\multirow{2}{*}{$\Delta
(3/2,3/2^-)$}} & 1 & 1.063  $\pm$  0.017  & 8.43/4  & 4 - 9 & \multicolumn{1}{c}{\multirow{2}{*}{3}} \\
\multicolumn{1}{c}{ } & 2  & 1.540  $\pm$  0.044  & 0.51/3  & 4 - 8 & & &\multicolumn{1}{c}{ } & 2  & 1.490  $\pm$  0.040  & 1.85/3  & 4 - 8 &\\
\multicolumn{1}{l}{\multirow{2}{*}{$\Delta (3/2,3/2^-)$}} & 1 & 1.089  $\pm$  0.014  & 8.76/5  & 4 - 10 & \multicolumn{1}{c}{\multirow{2}{*}{3}} & & \multicolumn{1}{l}{\multirow{2}{*}{$\Delta
(3/2,3/2^-)$}} & 1 & 1.056  $\pm$  0.018  & 10.25/4  & 4 - 9 & \multicolumn{1}{c}{\multirow{2}{*}{3}} \\
\multicolumn{1}{c}{ } & 2  & 1.543  $\pm$  0.044  & 0.76/3  & 4 - 8 & & &\multicolumn{1}{c}{ } & 2  & 1.482  $\pm$  0.044  & 1.95/3  & 4 - 8 &\\
\hline
\hline
\end{tabular}
\renewcommand{\arraystretch}{1}
\end{center}
\caption{Results of fits to the eigenvalues at the truncation levels $k = 0,16,20,30$ for $J = \frac{1}{2},\; \frac{3}{2}$ baryons. States are denoted by $n = 1, 2,...$. 
Corresponding mass values $am$ are given in lattice units; $t$
denotes the fit range and $\#i$ is the number of interpolators used in the construction of the cross-correlation 
matrix in a given quantum channel.}\label{Tablek0}
\end{table*} 

\end{appendix}

\section*{Erratum to the paper PRD 92, 074508 (2015) "Emergence of a new $SU(4)$
symmetry in the baryon spectrum"}
 
% \author{M.~Denissenya}
% %\email{mikhail.denissenya@uni-graz.at}
% \author{L.~Ya.~Glozman}
% %\email{leonid.glozman@uni-graz.at}
% \author{M.~Pak}
% %\email{markus.pak@uni-graz.at}
% \affiliation{Institut f\"ur Physik, FB Theoretische Physik, Universit\"at Graz, Universit\"atsplatz 5,
% 8010 Graz, Austria}
% 
% \begin{abstract}
% 
% \end{abstract}
% %\pacs{12.38Gc,11.25.-w,11.30.Rd}
% %\keywords{QCD \sep Chiral symmetry \sep Confinement \sep Lattice QCD \sep %Baryons}
% 
% \maketitle
In the article PRD 92, 074508 (2015) we have demonstrated a $SU(4)$
symmetry in the baryon spectrum upon elimination  of the quasi-zero modes of
the Dirac operator from the quark propagators. We have also claimed a
degeneracy of states belonging to different irreducible representations
of $SU(4)$. The latter statement is actually incorrect and the present lattice data
indicate existence of $SU(4)$ only. Whether a higher symmetry exists or
not cannot be deduced from these results.

In calculations we have used the following interpolator for $J=1/2$ $\Delta$-baryons:

$$ \Delta(\frac{3}{2},{\frac{1}{2}}^\pm) = i\varepsilon^{abc} \mathcal{P}_{\pm} 
 \gamma_i \gamma_5
 u^a  [u^{b T} C \gamma_i u^{c}]\; .
$$

\noindent
We have also discussed in the text of the paper another one

$$
 i\varepsilon^{abc} \mathcal{P}_{\pm} 
 \gamma_0 \gamma_5
 u^a [ u^{b T} C \gamma_0 u^{c}]\; .
$$

\noindent
We considered the former interpolator as to be independent from the
latter. This is actually incorrect, because their sum vanishes after performing of the Fierz transformation. Consequently, the interpolator 
$\Delta(\frac{3}{2},{\frac{1}{2}}^\pm)$ together with the interpolator (6)
of the paper
form a $(1,1/2)+(1/2,1)$ representation of $SU(2)_L \times SU(2)_R$. This
means that all three $N(\frac{1}{2},{\frac{1}{2}}^\pm)$ interpolators with
$\Delta(\frac{3}{2},{\frac{1}{2}}^\pm)$ form an irreducible dim=20 representation of $SU(4)$.

\end{document}